\documentclass[reprint,amsmath,amssymb,aps]{revtex4-1}

\usepackage{graphicx}
\usepackage{dcolumn}
\usepackage{bm}
\usepackage{hyperref}
\usepackage{epstopdf}
\usepackage{float}
\makeatletter\renewcommand{\maketag@@@}[1]{\hbox{\m@th\normalsize\normalfont#1}}
\usepackage{color, soul}

\begin{document}

\preprint{APS/123-QED}

\title{Tailoring Accelerating Beams in Phase Space}

\author{Yuanhui Wen, Yujie Chen$^{*}$, Yanfeng Zhang, Hui Chen, and Siyuan Yu}

\affiliation{State Key Laboratory of Optoelectronic Materials and Technologies, School of Electronics and Information Technology, Sun Yat-sen University, Guangzhou 510275, China
\\$^*$Corresponding author: chenyj69@mail.sysu.edu.cn}

\begin{abstract}
An appropriate design of wavefront will enable light fields propagating along arbitrary trajectories thus forming accelerating beams in free space. Previous ways of designing such accelerating beams mainly rely on caustic methods, which start from diffraction integrals and only deal with two-dimensional fields. Here we introduce a new perspective to construct accelerating beams in phase space by designing the corresponding Wigner distribution function (WDF). We find such a WDF-based method is capable of providing both the initial field distribution and the angular spectrum in need by projecting the WDF into the real space and the Fourier space respectively. Moreover, this approach applies to the construction of both two- and three-dimensional fields, greatly generalizing previous caustic methods. It may therefore open up a new route to construct highly-tailored accelerating beams and facilitate applications ranging from particle manipulation and trapping to optical routing as well as material processing.
\end{abstract}

\maketitle

\section{Introduction}

Wavefront contains the essential information of light including phase, amplitude and polarization, which can be engineered by various optical elements ranging from conventionally bulky to newly planar optical components \cite{Gansel1513,Yu333,Lin298,PhysRevLett.109.203903,Maguid1202,Nat.Mater}. An appropriate design of wavefront can lead to light fields capable of propagating along curved trajectories in free space, namely, accelerating beams. This seemingly counter-intuitive discovery was firstly revealed in 2007, in which light field with Airy distribution propagates along a parabolic trajectory \cite{Siviloglou:07,PhysRevLett.99.213901}. This peculiar property makes accelerating beams attractive to a variety of potential applications, such as micro-manipulation \cite{Nat.Photon.2008.201,Zhao2015Curved}, micro-machining \cite{apl}, imaging \cite{Nat.PhotonJS,Nat.Meth}, optical routing \cite{aplR}, light bullet \cite{chong2010airy,PhysRevLett.105.253901}, laser-assisted guiding of electric discharge \cite{Clericie1400111}, and plasma generation \cite{Polynkin229}. Moreover, due to the similar form of wave equations, the fundamental research of optical accelerating beams can be readily generalized to acoustic waves \cite{Nat.Common}, electron waves \cite{Nature} and plasmonic waves \cite{PhysRevLett.107.116802,PhysRevLett.107.126804,PhysRevLett.112.023903}, with broad and important influence beyond optics.

To make the applications flexible enough, accelerating beams propagating along various trajectories are required. Thus how to associate a desired propagating trajectory with an appropriate wavefront remains a crucial problem. Apart from keep finding other rare analytical solutions of the wave equation, a more efficient way is based on the caustic method, which associates the desired trajectory with an optical caustic \cite{Berry1980257}. This method was firstly implemented in real space \cite{PhysRevLett.106.213902,Froehly:11} and successfully realized arbitrary convex propagating trajectories for the two-dimensional (2D) light fields, while the caustic method in Fourier space was also proposed subsequently \cite{PhysRevA.88.043809,Bongiovanni2015Efficient,2016arXiv160700450W}. Therefore, for one predesigned trajectory, accelerating beams can be constructed by the caustic methods either in real space or in Fourier space, but the relationship between them is not clear yet. Moreover, the present caustic methods mainly focus on 2D fields, and the usual extension to the three-dimensional (3D) fields relies on the separability of the two transverse directions $(x,y)$ or rotation of the established 2D fields. Hence, to our knowledge, the real 3D fields, of which a typical one is a helicon wave, have not yet been investigated with present caustic methods.

In this Letter, we introduce a new perspective to describe and construct accelerating beams under the paraxial approximation. In contrast to previous caustic methods starting from the diffraction integrals in real space or Fourier space, we begin with the design of WDF in phase space \cite{Bastiaans:79,Nat.Photon.2012}. As WDF connects the space and spatial frequency domains, this new perspective is capable of revealing the relationship between the aforementioned caustic methods in real space and Fourier space in the 2D case. Additionally, in the 3D case beyond the analysis scope of previous caustic methods, our approach is still applicable and explicitly points out a new class of 3D accelerating beams, which includes the familiar helicon waves \cite{Alonzo:05,Daria:11,ADOM:ADOM201400315}.

\section{Theorectical Analysis}

The essential idea of our theoretical analysis is more generic and applicable to both 2D and 3D fields. For the sake of clarity, here we present the study in the 2D case, where the transverse variation is only along $X$-axis (independent of $Y$-axis) and the propagating direction is along $Z$-axis. Particularly, the coordinate for the initial plane $(X,Y,0)$ is denoted as $(x,y)$ for simplicity. As mentioned before, we begin with the phase-space design of WDF in the initial plane, which is defined as
\begin{small}
\begin{equation}
\begin{split}
W\left( {x,{k_x}} \right) & = \int {E\left( {x + {{x'} \mathord{\left/
 {\vphantom {{x'} 2}} \right.
 \kern-\nulldelimiterspace} 2}} \right){E^*}\left( {x - {{x'} \mathord{\left/
 {\vphantom {{x'} 2}} \right.
 \kern-\nulldelimiterspace} 2}} \right){e^{ - i{k_x}x'}}dx'} \\
& = \int {A\left( {{k_x} + {{{k_x^\prime}} \mathord{\left/
 {\vphantom {{{{k'}_x}} 2}} \right.
 \kern-\nulldelimiterspace} 2}} \right){A^*}\left( {{k_x} - {{{k_x^\prime}} \mathord{\left/
 {\vphantom {{{{k'}_x}} 2}} \right.
 \kern-\nulldelimiterspace} 2}} \right){e^{i{k_x^\prime}x}}d{k_x^\prime}}, 
\end{split}
\label{eq:1}
\end{equation}
\end{small}

\noindent where $E(x)$ is the initial field distribution, while $A(k_x)$ is the corresponding angular spectrum with $k_x$ to be the spatial frequency, and an asterisk denotes throughout complex conjugation. The physical meaning of WDF in optics can be approximately interpreted as the intensity of a light ray passing through the position $x$ with direction (or spatial frequency) $k_x$, but it may take a negative value due to the interference of waves \cite{Bastiaans:79,Nat.Photon.2012}. 

\begin{figure}
{\centerline{\includegraphics[width=8.5cm]{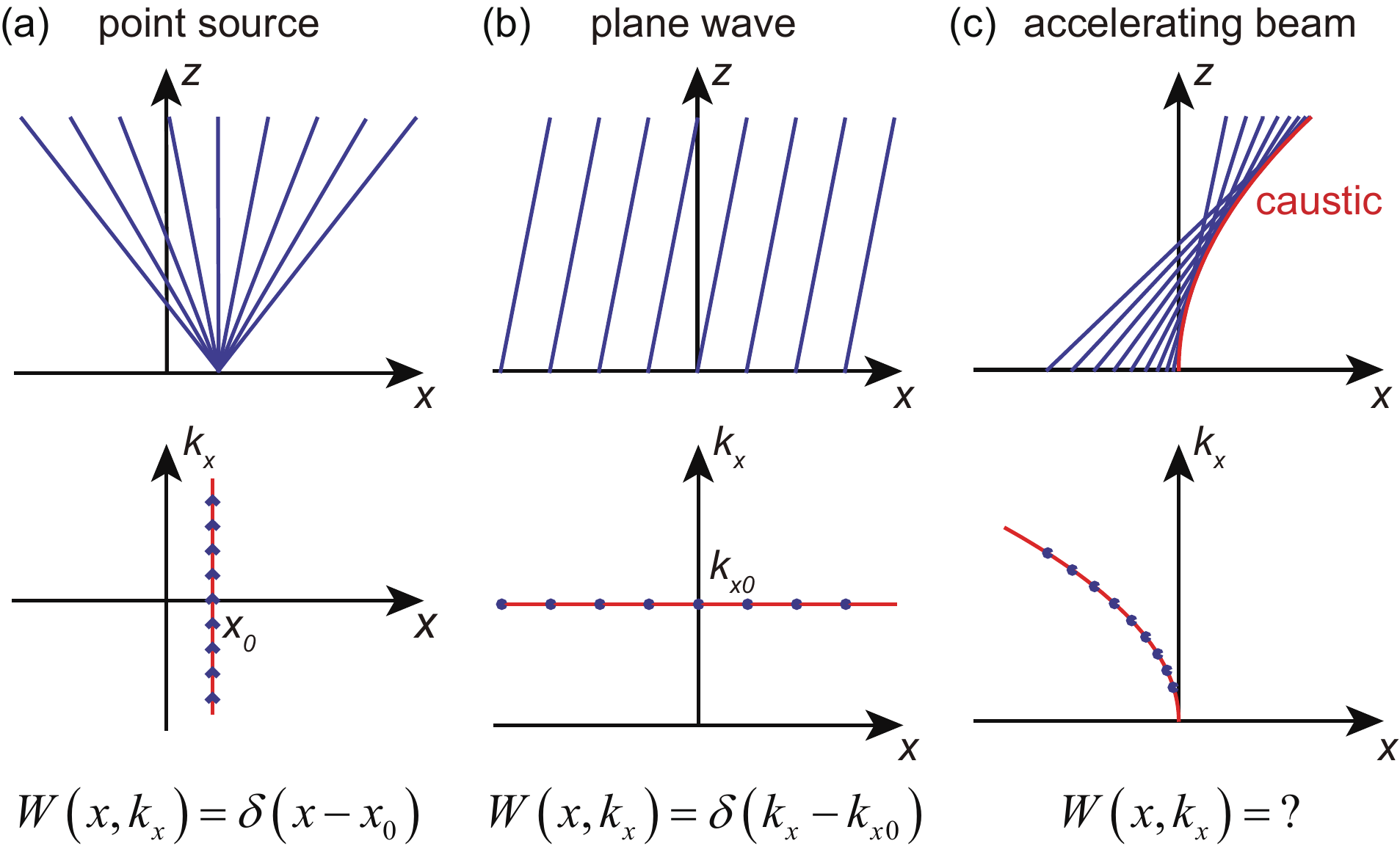}}}
\caption{\label{fig1}Schematic showing the light-ray picture and the corresponding form of WDF in phase space for (a) a point source, (b) a plane wave and (c) an accelerating beam.}
\end{figure}

In general, the WDF of a light field in phase space does not have an analytical form, except for some specific cases, such as a point source and a plane wave as shown in Figs.~\ref{fig1}(a) and \ref{fig1}(b). The WDFs of a point source located at the position $x_0$ and a plane wave with the spatial frequency $k_{x0}$ are $\delta \left( {x - {x_0}} \right)$ and $\delta \left( {{k_x} - {k_{x0}}} \right)$, respectively. The expressions have a clear physical picture that a point source contains the light rays emanating from the same initial point $x_0$ with all the directions $k_x$, while a plane wave contains the light rays emanating from all the initial points $x$ with the same direction $k_{x0}$. When it comes to the accelerating beam represented by its caustic trajectory, it is natural to ask what its WDF looks like in phase space and what is the relationship between the caustic trajectory and the WDF.

Note that it is impossible to figure out the exact WDF if only given the caustic trajectory, but we may construct a WDF with the inspiration from the WDF forms of a point source or a plane wave. Figure ~\ref{fig1}(c) shows that for a given caustic trajectory ${\rm{X = }}f\left( Z \right)$, each ray tangent to the caustic has a specific initial position $x{\rm{ = }}f\left( Z \right) - Zf'\left( Z \right) \buildrel \Delta \over = {F_1}\left( Z \right)$ and direction ${k_x}{\rm{ = }}kf'\left( Z \right) \buildrel \Delta \over = {F_2}\left( Z \right)$ (here $k$ is the wave number in free space), corresponding to a point $(x, k_x)$ located in phase space. Thus a WDF can be constructed by integrating all the above points and expressed mathematically as
\begin{small}
\begin{equation}
W\left( {x,{k_x}} \right) = \int_0^{{Z_{\max }}} {\delta \left[ {x - {F_1}\left( Z \right)} \right] \cdot \delta \left[ {{k_x} - {F_2}\left( Z \right)} \right]dZ}.  
\label{eq:2}
\end{equation}
\end{small} 

\noindent Based on this constructed WDF and its properties, we can directly obtain the initial field distribution $E\left( x \right){\rm{ = }}\rho \left( x \right){e^{i\varphi \left( x \right)}}$ by
\begin{small}
\begin{equation}
{\rho ^2}\left( x \right) = \int {W\left( {x,{k_x}} \right)d{k_x}},\ \varphi '\left( x \right) = \frac{{\int {{k_x}W\left( {x,{k_x}} \right)d{k_x}} }}{{\int {W\left( {x,{k_x}} \right)d{k_x}} }},
\label{eq:3}
\end{equation}
\end{small}

\noindent and the angular spectrum $A\left( {{k_x}} \right) = {\rm P}\left( {{k_x}} \right){e^{i\Phi \left( {{k_x}} \right)}}$ by
\begin{small}
\begin{equation}
{{\rm P}^2}\left( {{k_x}} \right) = \int {W\left( {x,{k_x}} \right)dx},\ \Phi '\left( {{k_x}} \right) =  - \frac{{\int {xW\left( {x,{k_x}} \right)dx} }}{{\int {W\left( {x,{k_x}} \right)dx} }}.
\label{eq:4}
\end{equation}
\end{small}

Therefore, once given the predesigned trajectory ${\rm{X = }}f\left( Z \right)$, a WDF can be constructed based on Eq.~(\ref{eq:2}) and then the necessary initial field distribution as well as the angular spectrum can be readily figured out using Eqs.~(\ref{eq:3}) and (\ref{eq:4}). This is the essential idea of our theoretical analysis, which is consistent for both 2D and 3D fields as mentioned before. In the following, we will apply this approach to some specific cases to demonstrate its wide suitability.

\section{Design of light beams in 2D space}

Firstly, we discuss the case of 2D fields and mainly consider the caustic of a power-law trajectory $X = f\left( Z \right) = {a_n}{Z^n}\left( {{a_n} > 0,n \ne 1} \right)$, which is the case widely investigated with previous caustic methods \cite{PhysRevLett.106.213902,Froehly:11,PhysRevA.88.043809,Bongiovanni2015Efficient} and thus suitable for comparison between our approach and those caustic methods. In this case, the constructed WDF can be figured out based on Eq.~(\ref{eq:2}), which reads

\begin{small}
\begin{equation}
W\left( {x,{k_x}} \right) = \frac{{\delta \left( {{k_x} - n{a_n}k{{\left\{ {{{ - x} \mathord{\left/
 {\vphantom {{ - x} {\left[ {\left( {n - 1} \right){a_n}} \right]}}} \right.
 \kern-\nulldelimiterspace} {\left[ {\left( {n - 1} \right){a_n}} \right]}}} \right\}}^{{{1 - 1} \mathord{\left/
 {\vphantom {{1 - 1} n}} \right.
 \kern-\nulldelimiterspace} n}}}} \right)}}{{\left| {n\left( {n - 1} \right){a_n}k{{\left\{ {{{ - x} \mathord{\left/
 {\vphantom {{ - x} {\left[ {\left( {n - 1} \right){a_n}} \right]}}} \right.
 \kern-\nulldelimiterspace} {\left[ {\left( {n - 1} \right){a_n}} \right]}}} \right\}}^{{{1 - 1} \mathord{\left/
 {\vphantom {{1 - 1} n}} \right.
 \kern-\nulldelimiterspace} n}}}} \right|}}.
\label{eq:5}
\end{equation}
\end{small}

\noindent This result shows the relationship between $x$ and $k_x$ in phase space, as previously shown in Fig.~\ref{fig1}(c). After a further calculation based on Eqs.~(\ref{eq:3}) and (\ref{eq:4}) (see Appendix \ref{A}), we can obtain the initial field distribution

\begin{small}
\begin{equation}
{\rho ^2}\left( x \right) = {A_n}{\left( { - x} \right)^{-1 + {1 \mathord{\left/
 {\vphantom {1 n}} \right.
 \kern-\nulldelimiterspace} n}}},\ \varphi \left( x \right) = {B_n}{\left( { - x} \right)^{2 - {1 \mathord{\left/
 {\vphantom {1 n}} \right.
 \kern-\nulldelimiterspace} n}}},
\label{eq:6}
\end{equation}
\end{small}

\noindent and the angular spectrum

\begin{small}
\begin{equation}
{{\rm P}^2}\left( {{k_x}} \right) = {C_n}k_x^{1 - {1 \mathord{\left/
 {\vphantom {1 {\left( {n - 1} \right)}}} \right.
 \kern-\nulldelimiterspace} {\left( {n - 1} \right)}}},\ \Phi \left( {{k_x}} \right) = {D_n}k_x^{2 + {1 \mathord{\left/
 {\vphantom {1 {\left( {n - 1} \right)}}} \right.
 \kern-\nulldelimiterspace} {\left( {n - 1} \right)}}}.
\label{eq:7}
\end{equation}
\end{small}

\noindent The obtained phase distributions in Eqs.~(\ref{eq:6}) and (\ref{eq:7}) are consistent with previous caustic methods either in real space \cite{PhysRevLett.106.213902,Froehly:11} or in Fourier space \cite{PhysRevA.88.043809,Bongiovanni2015Efficient}, which shows on one hand the reliability of our approach, and on the other hand the inner connection between the two previous caustic methods in the perspective of WDF. Apart from this, our analysis also gives initial amplitude distribution, which relates to the intensity variation along the caustic \cite{schley2014loss}.

\begin{figure}
{\centerline{\includegraphics[width=8.5cm]{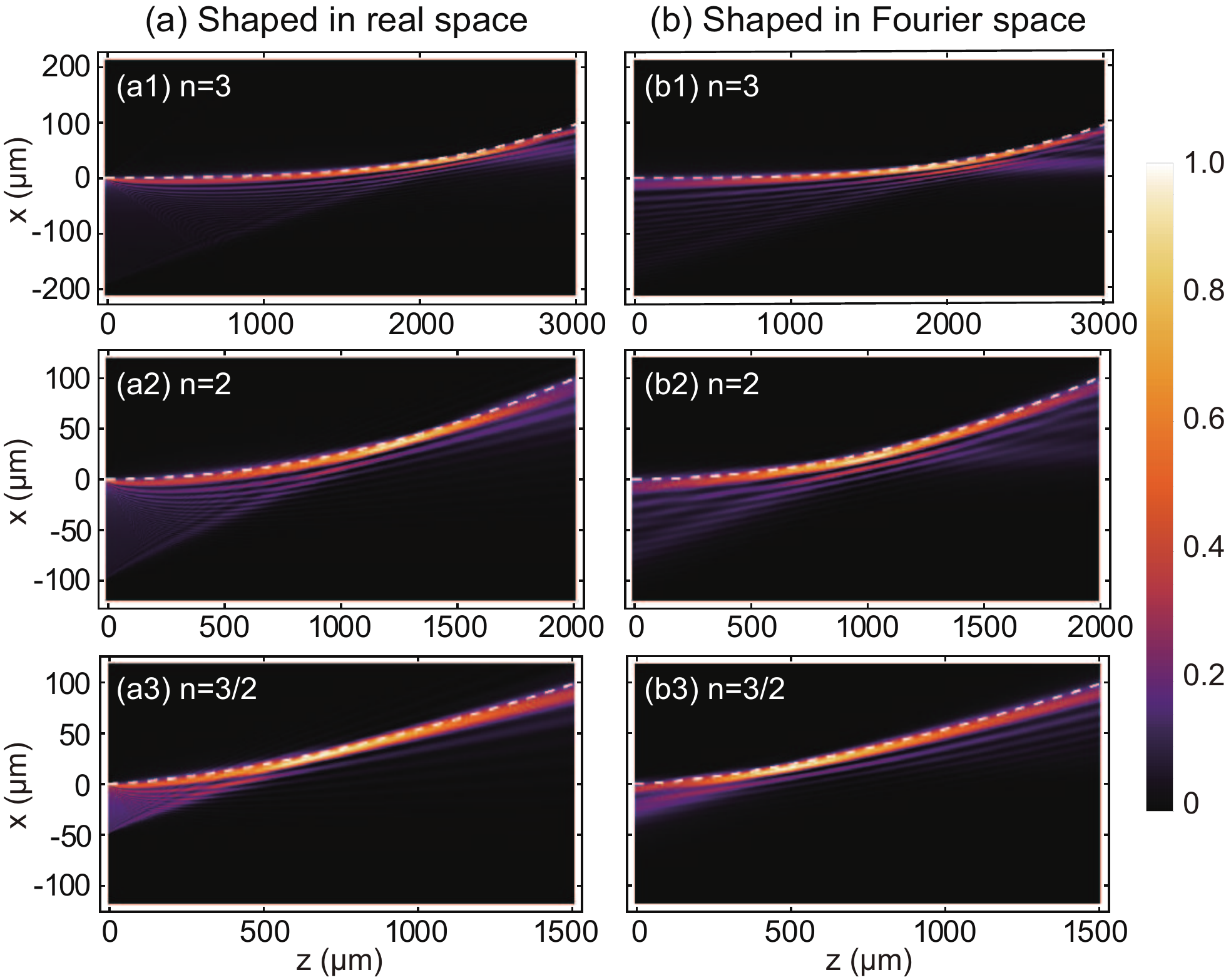}}}
\caption{\label{fig2}Intensity distribution of accelerating beams propagating along power-law trajectories with power $n$ chosen to be 3, 2, and $3/2$ for demonstration. The accelerating beams are constructed in (a) real space and (b) Fourier space based on the initial field distribution and the angular spectrum respectively. The predesigned trajectories are depicted out in dashed curves. }
\end{figure}

Based on the above theoretical results, some accelerating beams propagating along specific trajectories are demonstrated in Fig.~\ref{fig2}. The power $n$ of the trajectories include 3, 2, and $3/2$ with beams shaped in both real space and Fourier space. As shown clearly, the main lobes of the light beams propagate exactly along the predesigned trajectories, which further confirm the validity of our analysis. Moreover, for the same caustic trajectory, the beams shaped in real space and in Fourier space always behave similarly on the whole. This, on one hand, unifies the previous works' results obtained either in real space or in Fourier space \cite{PhysRevLett.106.213902,Froehly:11,PhysRevA.88.043809,Bongiovanni2015Efficient}, and on the other hand, provides more than one experimental generation scheme for the same caustic trajectory, including light field shaped in real space, Fourier space (realized by a Fourier lens), and even their combination.

\begin{figure}
{\centerline{\includegraphics[width=8.5cm]{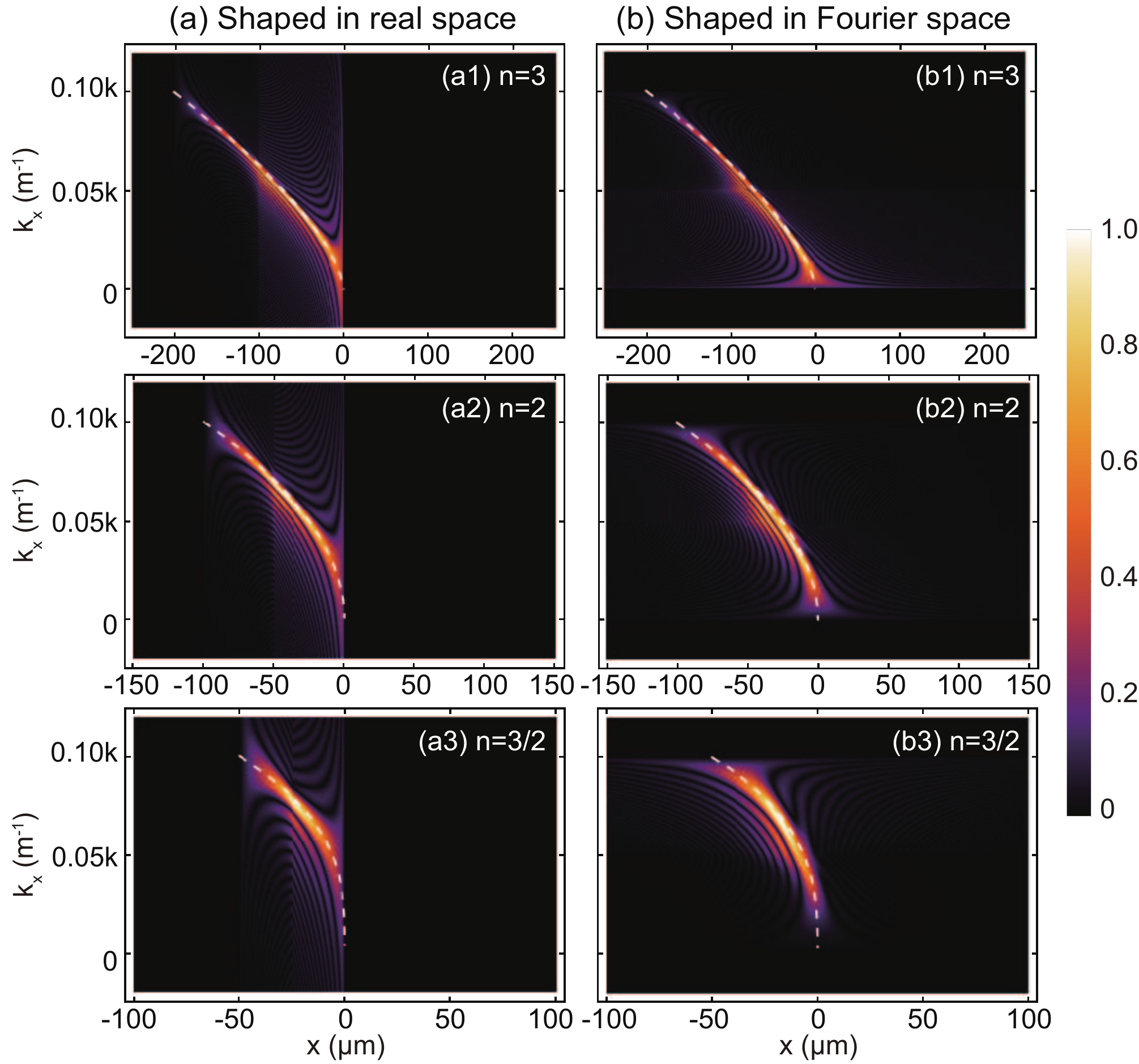}}}
\caption{\label{fig3}Comparison between the real WDFs and the construted WDFs for the above accelerating beams demonstrated in Fig.~\ref{fig2}. The real WDFs are calculated based on the definition in Eq.~(\ref{eq:1}) and then normalized and presented in absolute values here, while the constructed WDFs according to Eq.~(\ref{eq:2}) are presented with a corresponding section of dashed curves.}
\end{figure}

Furthermore, it is natural to wonder why this constructed WDF in phase space is capable of designing accelerating beams properly. To answer this question, the real WDFs in phase space of the above demonstrated accelerating beams are shown in Fig.~\ref{fig3} and for comparison, the constructed WDFs are depicted out with dashed curves. In fact, they are not identical, which means actually no light beams exactly correspond to these constructed WDFs. However, as can be seen in Fig.~\ref{fig3}, these constructed WDFs match well with the major part of the real ones, indicating that the constructed WDF is actually a simplified yet effective approximation to the real one by only considering the most significant light rays forming the caustic. Thus the phase-space design of WDF is a useful and effective approach in designing accelerating beams for specific applications.

\section{Design of light beams in 3D space}

It has been shown that the phase-space design of WDF is a useful tool to tailor accelerating beams along predesigned trajectory in 2D plane, which unifies previous caustic methods in real space and Fourier space. In the following, we will move to the case of 3D fields, including the situation where the light field along two transverse directions is inseparable and thus unable to be separated into the 2D case, beyond the analysis scope of the previous caustic methods. Here we firstly discuss a general 3D caustic curve in the form of 
 \begin{small}
\begin{equation}
X = f\left( {t + {t_0}} \right),\ Y = g\left( {t + {t_0}} \right),\ Z = at,
\label{eq:8}
\end{equation}
\end{small}

\noindent where $t$ is the parameter of this parametric equation, while $t_0$ and $a$ are undetermined variables. In this case, the constructed WDF can be expressed as
\begin{small}
\begin{equation}
W\left( {x,y,{k_x},{k_y}} \right) = \int_0^{{t_{\max }}} \begin{array}{l}
\delta \left[ {x - {F_1}\left( {t,{t_0}} \right),y - {G_1}\left( {t,{t_0}} \right)} \right] \cdot \\
\delta \left[ {{k_x} - {F_2}\left( {t,{t_0}} \right),{k_y} - {G_2}\left( {t,{t_0}} \right)} \right]dt
\end{array},       
\label{eq:9}
\end{equation}
\end{small}

\noindent where the expressions of ${F_1}\left( {t,{t_0}} \right)$, ${F_2}\left( {t,{t_0}} \right)$, ${G_1}\left( {t,{t_0}} \right)$, and ${G_2}\left( {t,{t_0}} \right)$ are given out in Appendix \ref{B}. After a further detailed derivation (see Appendix \ref{C}), it is interesting to arrive at a natural constraint for the trajectory

\begin{small}
\begin{equation}
{\left[ {f'\left( {t + {t_0}} \right)} \right]^2}{\rm{ + }}{\left[ {g'\left( {t + {t_0}} \right)} \right]^2}{\rm{ = const}} \buildrel \Delta \over = {u^2},
\label{eq:10}
\end{equation}
\end{small}

\noindent which requires $k_r$ to be a constant, corresponding to the requirement for nondiffracting beams \cite{PhysRevLett.105.013902}. Note that there is one class of caustic trajectories satisfying the above constraint 

\begin{small}
\begin{equation}
\begin{split}
f'\left( {t + {t_0}} \right) & {\rm{ =  - }}u \sin \left[ {h\left( {t + {t_0}} \right)} \right] \\
g'\left( {t + {t_0}} \right) & {\rm{ = }}u \cos \left[ {h\left( {t + {t_0}} \right)} \right]
\end{split}~,
\label{eq:11}
\end{equation}
\end{small}

\noindent with $h\left( {t + {t_0}} \right)$ to be an arbitrary function, and we will mainly discuss these trajectories in the following.

\begin{figure}
{\centerline{\includegraphics[width=8.5cm]{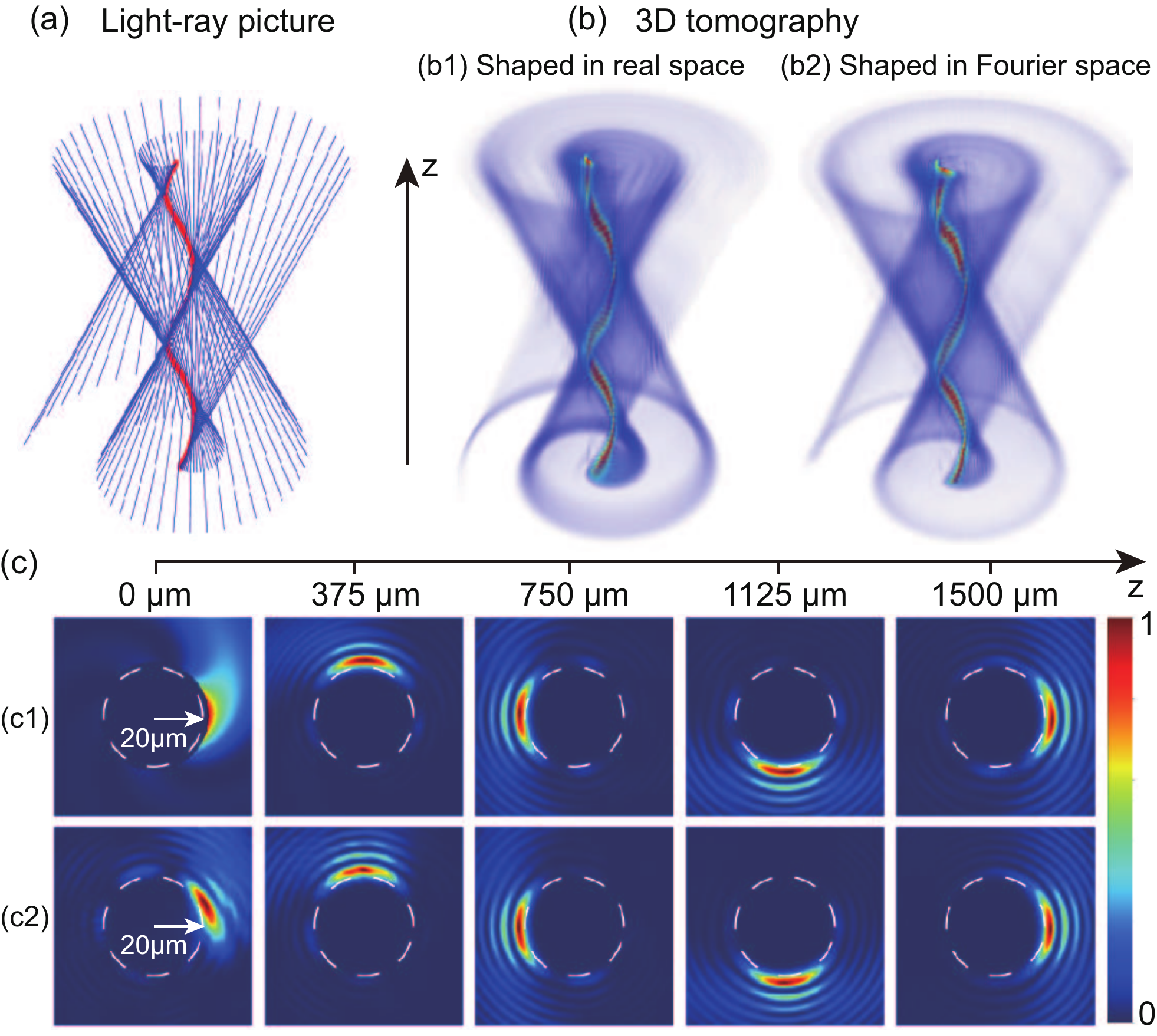}}}
\caption{\label{fig4}Simulation results of constructed helical beams. (a) A light-ray picture describing a bundle of light rays in blue forming a helical caustic in red. (b) The 3D tomography showing the propagation dynamics of two-cycle helical beams constructed in (b1) real space and (b2) Fourier space respectively. (c) The cross-sectional intensity distributions in the first cycle are presented in detail for the above helical beams shaped in (c1) real space and (c2) Fourier space, with a dashed circle in the radius of $20~\mu$m representing the projection of the helical caustic to the initial plane.}
\end{figure}

Considering the case of $h\left( {t + {t_0}} \right) = t + {t_0}$ corresponding to a helical trajectory as shown in Fig.~\ref{fig4}(a) with a red curve. Applying the constructed WDF in the same way, we can also obtained the required initial field distribution 

\begin{small}
\begin{equation}
\begin{split}
{\rho ^2}\left( {r,\theta } \right) & = \frac{{\delta \left[ {\theta  - \sqrt {{{{r^2}} \mathord{\left/
 {\vphantom {{{r^2}} {{u^2}}}} \right.
 \kern-\nulldelimiterspace} {{u^2}}} - 1}  + \arccos \left( {{u \mathord{\left/
 {\vphantom {u r}} \right.
 \kern-\nulldelimiterspace} r}} \right) - {t_0}} \right]}}{{\sqrt {{r^2} - {u^2}} }} \\
\varphi \left( {r,\theta } \right) & = {{\left[ {\theta  - \sqrt {{{{r^2}} \mathord{\left/
 {\vphantom {{{r^2}} {{u^2}}}} \right.
 \kern-\nulldelimiterspace} {{u^2}}} - 1}  + \arccos \left( {{u \mathord{\left/
 {\vphantom {u r}} \right.
 \kern-\nulldelimiterspace} r}} \right)} \right]{u^2}k} \mathord{\left/
 {\vphantom {{\left[ {\theta  - \sqrt {{{{r^2}} \mathord{\left/
 {\vphantom {{{r^2}} {{u^2}}}} \right.
 \kern-\nulldelimiterspace} {{u^2}}} - 1}  + \arccos \left( {{u \mathord{\left/
 {\vphantom {u r}} \right.
 \kern-\nulldelimiterspace} r}} \right)} \right]{u^2}k} a}} \right.
 \kern-\nulldelimiterspace} a}
\end{split}~,
\label{eq:12}
\end{equation}
\end{small}

\noindent and angular spectrum
\begin{small}
\begin{equation}
\begin{split}
{{\rm P}^2}\left( {{k_r},{k_\theta }} \right) & = {{\delta \left( {{k_r} - {{uk} \mathord{\left/
 {\vphantom {{uk} a}} \right.
 \kern-\nulldelimiterspace} a}} \right)} \mathord{\left/
 {\vphantom {{\delta \left( {{k_r} - {{uk} \mathord{\left/
 {\vphantom {{uk} a}} \right.
 \kern-\nulldelimiterspace} a}} \right)} {{k_r}}}} \right.
 \kern-\nulldelimiterspace} {{k_r}}} \\
\Phi \left( {{k_r},{k_\theta }} \right) & = u{k_r}\left( {{k_\theta } - {\pi  \mathord{\left/
 {\vphantom {\pi  {2 - {t_0}}}} \right.
 \kern-\nulldelimiterspace} {2 - {t_0}}}} \right)
\end{split}~,
\label{eq:13}
\end{equation}
\end{small}

\noindent in the polar coordinate.

It is noted that the intensity of the initial field distribution and the angular spectrum are both Dirac functions, which results from the predesigned caustic being a curve rather than a surface in 3D space, as shown in Fig.~\ref{fig4}(a). If the caustic curve slightly extends to a surface (e.g. $t_0$ or $a$ varies within a certain range in this case), the intensity distribution in the form of Dirac function will be replaced by a finite-width function, and the choice of this width needs to balance the confinement and divergence of the main lobe. 

On the other hand, the phase distribution of the angular spectrum is similar to the one directly proposed for helico-conical beams \cite{Alonzo:05,Daria:11}. Although it has been observed in these works that the head of the spiral propagates along a helical trajectory around the optical axis, no further explanation is given for this interesting phenomenon. Here we start from our constructed WDF and naturally associate a helical trajectory with such a phase distribution in Fourier space. Moreover, we also give out its counterpart constructed in real space, which is similar to that of Ref. \cite{ADOM:ADOM201400315}. Here we investigate in the paraxial regime and provide a more clear relation between the helical caustic and the initial spiral-shape field distribution.

\begin{figure}
{\centerline{\includegraphics[width=8.5cm]{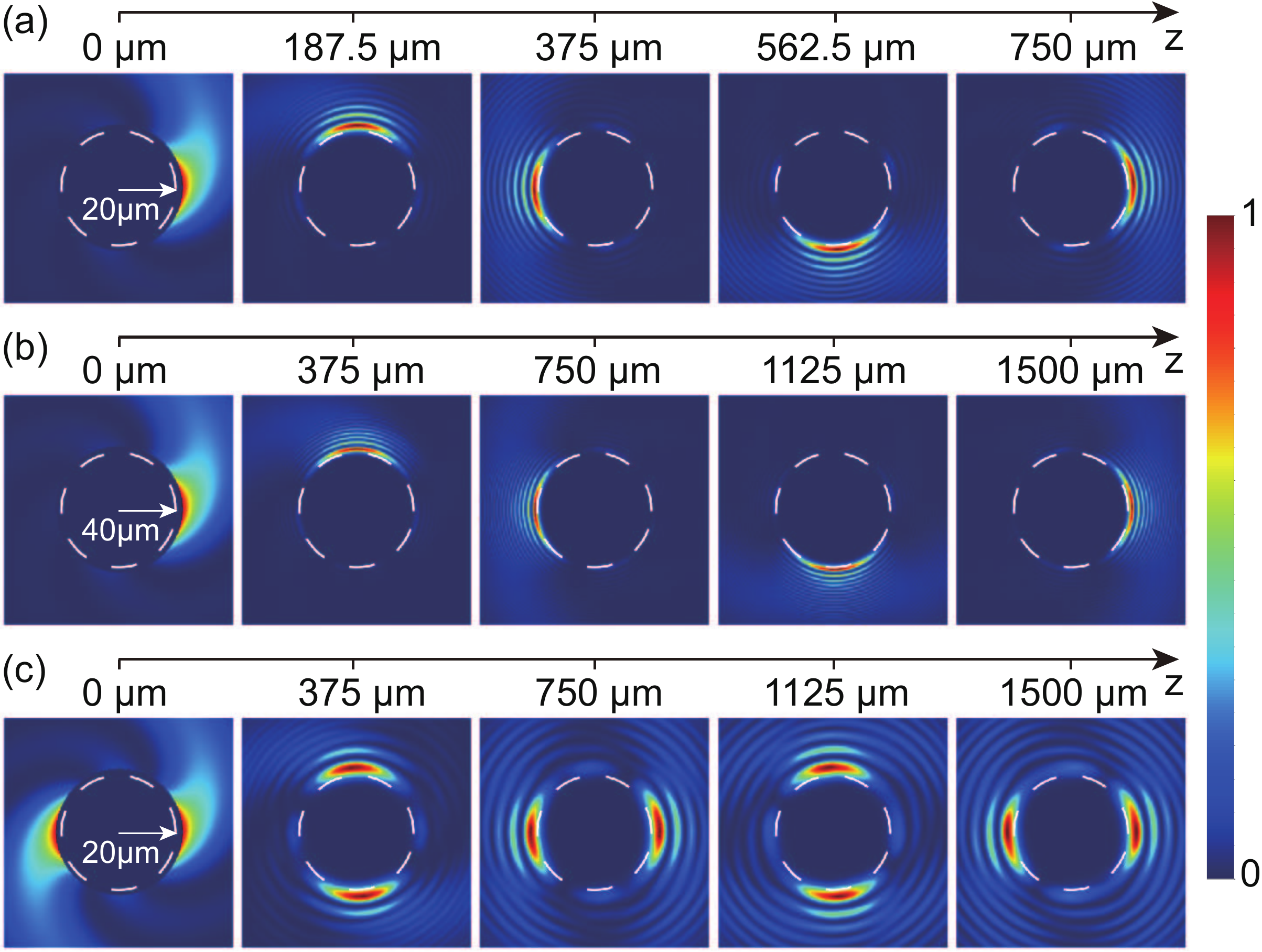}}}
\caption{\label{fig5}The cross-sectional intensity distributions of helical beams constructed in real space under different parameters in comparison with Fig.~\ref{fig4}(c1). (a) radius $u = 20~\mu m$, peroid $2\pi a = 750~\mu m$, (b) radius $u = 40~\mu m$, period $2\pi a = 1500~\mu m$, (c) raduis $u = 20~\mu m$, period $2\pi a = 1500~\mu m$, two lobes.}
\end{figure}

Based on Eqs.~(\ref{eq:12}) and (\ref{eq:13}), we have designed a two-cycle helical caustic with the radius $u = 20~\mu m$ and the period $2\pi a = 1500~\mu m$ for demonstration. The simulation results are shown in Fig.~\ref{fig4}(b) in the form of 3D tomography to illustrate the propagation dynamics, while the cross-sectional intensity distributions in the first cycle are presented in Fig.~\ref{fig4}(c), with a dashed circle representing the projection of the predesigned helical caustic to the initial plane. It is worth mentioning that this constructed helical beam is highly adjustable and customizable, including the radius, the period, the number of periods and even the number of main lobes as shown in Fig.~\ref{fig5}, which will provide great flexibility in applications.

\begin{figure}
{\centerline{\includegraphics[width=8.5cm]{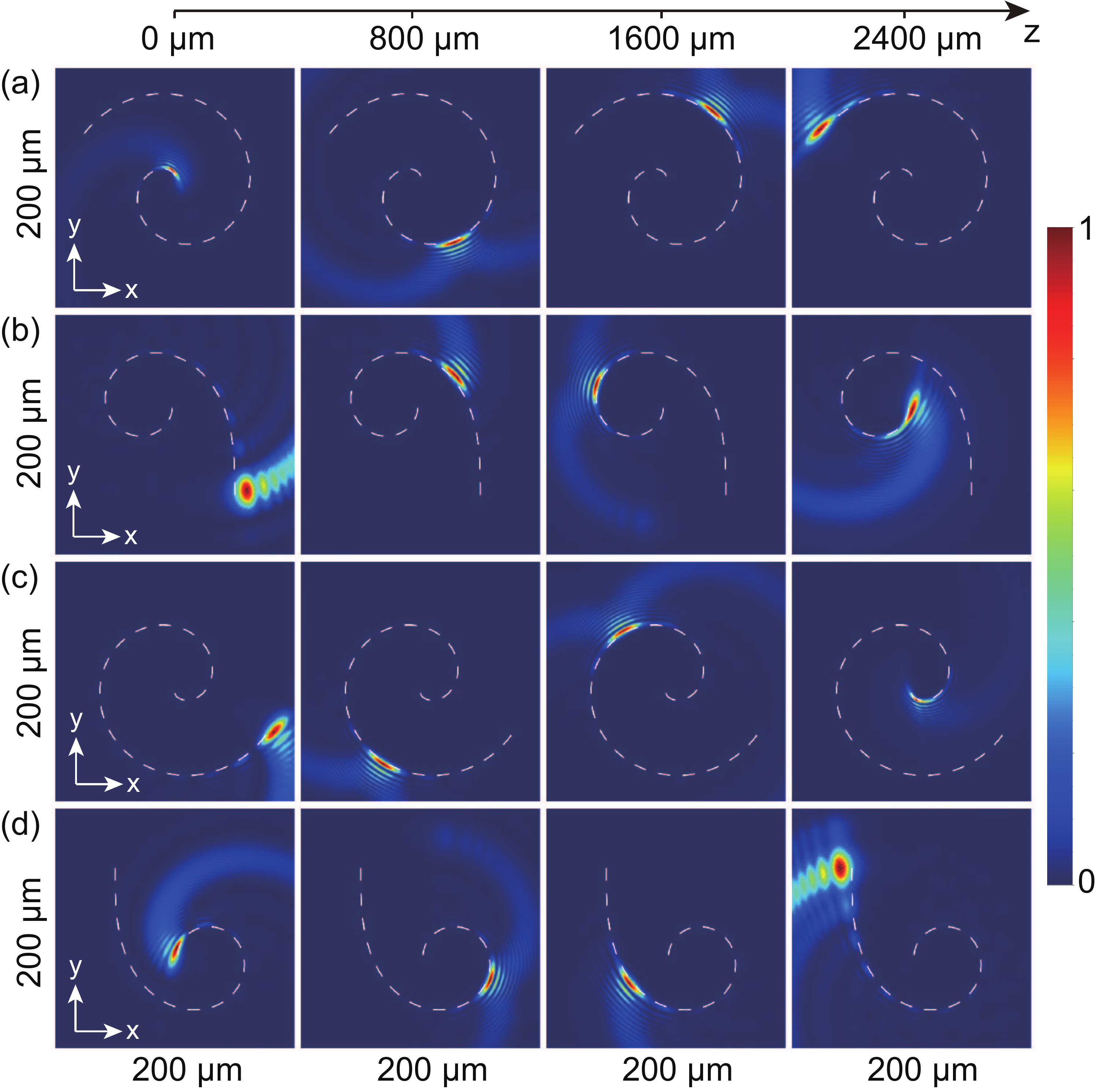}}}
\caption{\label{fig6}Simulating propagation dynamics for the case of (a)  $h\left( {t + {t_0}} \right) = {\left( {t + {t_0}} \right)^{1/2}}$ and (b) $h\left( {t + {t_0}} \right) = {\left( {t + {t_0}} \right)^2}$.
The cross-sectional intensity distribution during propagation is presented for the accelerating beams constructed in Fourier space and the dash curve in it is (a) an involute of circle, and (b) a Cornu spiral, representing the projection of the predesigned caustic into the initial plane.}
\end{figure}

Apart from the case of $h\left( {t + {t_0}} \right) = t + {t_0}$ corresponding to a helical caustic, there are still a variety of 3D caustic contained in Eq.~(\ref{eq:10}). In the following, we will discuss another two cases of $h\left( {t + {t_0}} \right) = {\left( {t + {t_0}} \right)^{1/2}}$ and $h\left( {t + {t_0}} \right) = {\left( {t + {t_0}} \right)^2}$. The acclerating beams are constructed in Fourier space and their propagation dynamics are presented in Fig.~\ref{fig6}(a) and \ref{fig6}(b). The dash curve in the cross-sectional intensity distribution is the projection of the predesigned caustic to the initial plane, which is an involute of circle for the case of $h\left( {t + {t_0}} \right) = {\left( {t + {t_0}} \right)^{1/2}}$ and an Cornu spiral for $h\left( {t + {t_0}} \right) = {\left( {t + {t_0}} \right)^2}$. As can be seen in Fig.~\ref{fig6}(a) and \ref{fig6}(b), the main lobe of each beam propagates exactly along the predesigned trajectory and are well confined in it. When it comes to the end of the trajectory, the main lobe starts to spread out as expected. In addition, the direction of these beams' rotation can also be reversed, corresponding to the cases of $h\left( {t + {t_0}} \right) = -{\left( {t + {t_0}} \right)^{1/2}}$ and $h\left( {t + {t_0}} \right) = -{\left( {t + {t_0}} \right)^2}$ as shown in Fig.~\ref{fig6}(c) and \ref{fig6}(d). These results further confirm the existence of a wide-ranging class of 3D accelerating beams contained in Eq.~(\ref{eq:10}).

\section{Conclusion}

In conclusion, we propose a new approach to construct accelerating beams by designing the WDF in phase space. Based on the constructed WDF, we can readily obtain the necessary initial field distribution in real space and angular spectrum in Fourier space simultaneously. To show its wide suitability, firstly in the 2D case, we analyze accelerating beams moving along power-law trajectories, which unifies previous caustic methods proposed in real space and Fourier space; later in the 3D case, we reveal a new class of 3D accelerating beams including the familiar helical beam, beyond the analysis scope of previous caustic methods. The phase-space design of accelerating beams put forward here is intuitive, widely-applicable, and easy-handling, which is expected to find important applications in demand of highly tailored accelerating beams, ranging from particle manipulation to material processing. It also deepens our understanding on accelerating beams within a new perspective and may further advance the field of engineering wavefronts for future applications.

\section{Acknowledgements}

This work is supported by the National Basic Research Program of China (973 Program) (2014CB340000 and 2012CB315702), the Natural Science Foundations of China (61323001, 61490715, 51403244, and 11304401), and the Natural Science Foundation of Guangdong Province (2014A030313104).

\appendix

\section{Derivations for Eqs. (6) and (7) in the case of 2D space}

\label{A}

Based on Eqs.~(\ref{eq:3}) and (\ref{eq:4}) as well as the given WDF in Eq.~(\ref{eq:5}), we can calculate the initial field distribution
\begin{small}
\begin{equation}
\begin{split}
{\rho ^2}\left( x \right) & = \int {W\left( {x,{k_x}} \right)d{k_x}} \\ 
& = \int {\frac{{\delta \left( {{k_x} - n{a_n}k{{\left\{ {{{ - x} \mathord{\left/
 {\vphantom {{ - x} {\left[ {\left( {n - 1} \right){a_n}} \right]}}} \right.
 \kern-\nulldelimiterspace} {\left[ {\left( {n - 1} \right){a_n}} \right]}}} \right\}}^{1 - {1 \mathord{\left/
 {\vphantom {1 n}} \right.
 \kern-\nulldelimiterspace} n}}}} \right)}}{{n\left( {n - 1} \right){a_n}k{{\left\{ {{{ - x} \mathord{\left/
 {\vphantom {{ - x} {\left[ {\left( {n - 1} \right){a_n}} \right]}}} \right.
 \kern-\nulldelimiterspace} {\left[ {\left( {n - 1} \right){a_n}} \right]}}} \right\}}^{1 - {1 \mathord{\left/
 {\vphantom {1 n}} \right.
 \kern-\nulldelimiterspace} n}}}}}d{k_x}} \\
& = {\left( {n\left( {n - 1} \right){a_n}k{{\left\{ {{{ - x} \mathord{\left/
 {\vphantom {{ - x} {\left[ {\left( {n - 1} \right){a_n}} \right]}}} \right.
 \kern-\nulldelimiterspace} {\left[ {\left( {n - 1} \right){a_n}} \right]}}} \right\}}^{1 - {1 \mathord{\left/
 {\vphantom {1 n}} \right.
 \kern-\nulldelimiterspace} n}}}} \right)^{ - 1}} \\
& = {k^{ - 1}}{n^{ - 1}}{\left[ {\left( {n - 1} \right){a_n}} \right]^{ - {1 \mathord{\left/
 {\vphantom {1 n}} \right.
 \kern-\nulldelimiterspace} n}}}{\left( { - x} \right)^{ - 1 + {1 \mathord{\left/
 {\vphantom {1 n}} \right.
 \kern-\nulldelimiterspace} n}}} \\
& \buildrel \Delta \over = {A_n}{\left( { - x} \right)^{ - 1 + {1 \mathord{\left/
 {\vphantom {1 n}} \right.
 \kern-\nulldelimiterspace} n}}},
\end{split}
\label{eq:A1}
\end{equation}
\end{small}
\begin{small}
\begin{equation}
\begin{split}
\varphi '\left( x \right) & = \frac{{\int {{k_x}W\left( {x,{k_x}} \right)d{k_x}} }}{{\int {W\left( {x,{k_x}} \right)d{k_x}} }} \\
& = \frac{{\int {{k_x}\delta \left( {{k_x} - n{a_n}k{{\left\{ {{{ - x} \mathord{\left/
 {\vphantom {{ - x} {\left[ {\left( {n - 1} \right){a_n}} \right]}}} \right.
 \kern-\nulldelimiterspace} {\left[ {\left( {n - 1} \right){a_n}} \right]}}} \right\}}^{1 - {1 \mathord{\left/
 {\vphantom {1 n}} \right.
 \kern-\nulldelimiterspace} n}}}} \right)d{k_x}} }}{{\int {\delta \left( {{k_x} - n{a_n}k{{\left\{ {{{ - x} \mathord{\left/
 {\vphantom {{ - x} {\left[ {\left( {n - 1} \right){a_n}} \right]}}} \right.
 \kern-\nulldelimiterspace} {\left[ {\left( {n - 1} \right){a_n}} \right]}}} \right\}}^{1 - {1 \mathord{\left/
 {\vphantom {1 n}} \right.
 \kern-\nulldelimiterspace} n}}}} \right)d{k_x}} }} \\
& = n{a_n}k{\left\{ {{{ - x} \mathord{\left/
 {\vphantom {{ - x} {\left[ {\left( {n - 1} \right){a_n}} \right]}}} \right.
 \kern-\nulldelimiterspace} {\left[ {\left( {n - 1} \right){a_n}} \right]}}} \right\}^{1 - {1 \mathord{\left/
 {\vphantom {1 n}} \right.
 \kern-\nulldelimiterspace} n}}} \\
& = kn{\left( {n - 1} \right)^{{1 \mathord{\left/
 {\vphantom {1 n}} \right.
 \kern-\nulldelimiterspace} n} - 1}}a_n^{{1 \mathord{\left/
 {\vphantom {1 n}} \right.
 \kern-\nulldelimiterspace} n}}{\left( { - x} \right)^{1 - {1 \mathord{\left/
 {\vphantom {1 n}} \right.
 \kern-\nulldelimiterspace} n}}},
\end{split}
\label{eq:A2}
\end{equation}
\end{small}
\begin{small}
\begin{equation}
\begin{split}
\varphi \left( x \right) & =  - k{n^2}{\left( {n - 1} \right)^{{1 \mathord{\left/
 {\vphantom {1 n}} \right.
 \kern-\nulldelimiterspace} n} - 1}}{\left( {2n - 1} \right)^{ - 1}}a_n^{{1 \mathord{\left/
 {\vphantom {1 n}} \right.
 \kern-\nulldelimiterspace} n}}{\left( { - x} \right)^{2 - {1 \mathord{\left/
 {\vphantom {1 n}} \right.
 \kern-\nulldelimiterspace} n}}} \\ 
& \buildrel \Delta \over = {B_n}{\left( { - x} \right)^{2 - {1 \mathord{\left/
 {\vphantom {1 n}} \right.
 \kern-\nulldelimiterspace} n}}},
\end{split}
\label{eq:A3}
\end{equation}
\end{small}

\noindent as well as the angular spectrum
\begin{small}
\begin{equation}
\begin{split}
{{\rm P}^2}\left( {{k_x}} \right) & = \int {W\left( {x,{k_x}} \right)dx} \\ 
& = \int {\frac{{\delta \left( {{k_x} - n{a_n}k{{\left\{ {{{ - x} \mathord{\left/
 {\vphantom {{ - x} {\left[ {\left( {n - 1} \right){a_n}} \right]}}} \right.
 \kern-\nulldelimiterspace} {\left[ {\left( {n - 1} \right){a_n}} \right]}}} \right\}}^{1 - {1 \mathord{\left/
 {\vphantom {1 n}} \right.
 \kern-\nulldelimiterspace} n}}}} \right)}}{{n\left( {n - 1} \right){a_n}k{{\left\{ {{{ - x} \mathord{\left/
 {\vphantom {{ - x} {\left[ {\left( {n - 1} \right){a_n}} \right]}}} \right.
 \kern-\nulldelimiterspace} {\left[ {\left( {n - 1} \right){a_n}} \right]}}} \right\}}^{1 - {1 \mathord{\left/
 {\vphantom {1 n}} \right.
 \kern-\nulldelimiterspace} n}}}}}dx} \\ 
& = {\left\{ {n\left( {n - 1} \right){a_n}{{\left[ {{{{k_x}} \mathord{\left/
 {\vphantom {{{k_x}} {\left( {n{a_n}k} \right)}}} \right.
 \kern-\nulldelimiterspace} {\left( {n{a_n}k} \right)}}} \right]}^{1 - {1 \mathord{\left/
 {\vphantom {1 {\left( {n - 1} \right)}}} \right.
 \kern-\nulldelimiterspace} {\left( {n - 1} \right)}}}}} \right\}^{ - 1}} \\
& = {k^{{1 \mathord{\left/
 {\vphantom {1 {\left( {n - 1} \right) - 1}}} \right.
 \kern-\nulldelimiterspace} {\left( {n - 1} \right) - 1}}}}\left( {n - 1} \right){\left( {n{a_n}} \right)^{{1 \mathord{\left/
 {\vphantom {1 {\left( {n - 1} \right)}}} \right.
 \kern-\nulldelimiterspace} {\left( {n - 1} \right)}}}}k_x^{1 - {1 \mathord{\left/
 {\vphantom {1 {\left( {n - 1} \right)}}} \right.
 \kern-\nulldelimiterspace} {\left( {n - 1} \right)}}} \\
& \buildrel \Delta \over = {C_n}k_x^{1 - {1 \mathord{\left/
 {\vphantom {1 {\left( {n - 1} \right)}}} \right.
 \kern-\nulldelimiterspace} {\left( {n - 1} \right)}}},
\end{split}
\label{eq:A4}
\end{equation}
\end{small}
\begin{small}
\begin{equation}
\begin{split}
\Phi '\left( {{k_x}} \right) & =  - \frac{{\int {xW\left( {x,{k_x}} \right)dx} }}{{\int {W\left( {x,{k_x}} \right)dx} }} \\
& =  - \frac{{\int {x\delta \left\{ {x + \left( {n - 1} \right){a_n}{{\left[ {{{{k_x}} \mathord{\left/
 {\vphantom {{{k_x}} {\left( {n{a_n}k} \right)}}} \right.
 \kern-\nulldelimiterspace} {\left( {n{a_n}k} \right)}}} \right]}^{1 + {1 \mathord{\left/
 {\vphantom {1 {\left( {n - 1} \right)}}} \right.
 \kern-\nulldelimiterspace} {\left( {n - 1} \right)}}}}} \right\}dx} }}{{\int {\delta \left\{ {x + \left( {n - 1} \right){a_n}{{\left[ {{{{k_x}} \mathord{\left/
 {\vphantom {{{k_x}} {\left( {n{a_n}k} \right)}}} \right.
 \kern-\nulldelimiterspace} {\left( {n{a_n}k} \right)}}} \right]}^{1 + {1 \mathord{\left/
 {\vphantom {1 {\left( {n - 1} \right)}}} \right.
 \kern-\nulldelimiterspace} {\left( {n - 1} \right)}}}}} \right\}dx} }} \\
& = \left( {n - 1} \right){a_n}{\left[ {{{{k_x}} \mathord{\left/
 {\vphantom {{{k_x}} {\left( {n{a_n}k} \right)}}} \right.
 \kern-\nulldelimiterspace} {\left( {n{a_n}k} \right)}}} \right]^{1 + {1 \mathord{\left/
 {\vphantom {1 {\left( {n - 1} \right)}}} \right.
 \kern-\nulldelimiterspace} {\left( {n - 1} \right)}}}} \\
& = {\left( {kn} \right)^{ - 1 - {1 \mathord{\left/
 {\vphantom {1 {\left( {n - 1} \right)}}} \right.
 \kern-\nulldelimiterspace} {\left( {n - 1} \right)}}}}\left( {n - 1} \right)a_n^{ - {1 \mathord{\left/
 {\vphantom {1 {\left( {n - 1} \right)}}} \right.
 \kern-\nulldelimiterspace} {\left( {n - 1} \right)}}}k_x^{1 + {1 \mathord{\left/
 {\vphantom {1 {\left( {n - 1} \right)}}} \right.
 \kern-\nulldelimiterspace} {\left( {n - 1} \right)}}},
\end{split}
\label{eq:A5}
\end{equation}
\end{small}
\begin{small}
\begin{equation}
\begin{split}
\Phi \left( {{k_x}} \right) & = {\left( {kn} \right)^{ - 1 - {1 \mathord{\left/
 {\vphantom {1 {\left( {n - 1} \right)}}} \right.
 \kern-\nulldelimiterspace} {\left( {n - 1} \right)}}}}{\left( {n - 1} \right)^2}{\left( {2n - 1} \right)^{ - 1}}a_n^{ - {1 \mathord{\left/
 {\vphantom {1 {\left( {n - 1} \right)}}} \right.
 \kern-\nulldelimiterspace} {\left( {n - 1} \right)}}}k_x^{2 + {1 \mathord{\left/
 {\vphantom {1 {\left( {n - 1} \right)}}} \right.
 \kern-\nulldelimiterspace} {\left( {n - 1} \right)}}} \\ 
& \buildrel \Delta \over = {D_n}k_x^{2 + {1 \mathord{\left/
 {\vphantom {1 {\left( {n - 1} \right)}}} \right.
 \kern-\nulldelimiterspace} {\left( {n - 1} \right)}}}.
\end{split}
\label{eq:A6}
\end{equation}
\end{small}

\section{Expressions for some symbols }

\label{B}

\begin{small}
\begin{equation}
{F_1}\left( {t,{t_0}} \right) = f\left( {t + {t_0}} \right) - tf'\left( {t + {t_0}} \right),
\label{eq:B1}
\end{equation}
\end{small}
\begin{small}
\begin{equation}
{F_2}\left( {t,{t_0}} \right) = k{a^{ - 1}}f'\left( {t + {t_0}} \right),
\label{eq:B2}
\end{equation}
\end{small}
\begin{small}
\begin{equation}
{G_1}\left( {t,{t_0}} \right) = g\left( {t + {t_0}} \right) - tg'\left( {t + {t_0}} \right),
\label{eq:B3}
\end{equation}
\end{small}
\begin{small}
\begin{equation}
{G_2}\left( {t,{t_0}} \right) = k{a^{ - 1}}g'\left( {t + {t_0}} \right).
\label{eq:B4}
\end{equation}
\end{small}

\section{Derivations for Eq.~(\ref{eq:10}), a natural constraint in the case of 3D space}

\label{C}

First considering the real space, based on Eq.~(\ref{eq:9}), we can calculate the initial field distribution including the intensity distribution
\begin{small}
\begin{equation}
\begin{split}
{\rho ^2}\left( {x,y} \right) & =\iint{{W\left( {x,y,{k_x},{k_y}} \right)d{k_x}d{k_y}}}\\  
& = \int_0^{{t_{\max }}} {\delta \left[ {x - {F_1}\left( {t,{t_0}} \right),y - {G_1}\left( {t,{t_0}} \right)} \right]dt}, 
\end{split}
\label{eq:C1}
\end{equation}
\end{small}

\noindent and the derivatives of phase distribution with respect to space coordinates $x$ and $y$
\begin{small}
\begin{equation}
\begin{split}
\frac{{\partial \varphi \left( {x,y} \right)}}{{\partial x}} & = \frac{{\iint{{{k_x}W\left( {x,y,{k_x},{k_y}} \right)d{k_x}d{k_y}}}}}{{\iint{{W\left( {x,y,{k_x},{k_y}} \right)d{k_x}d{k_y}}}}}\\
& = \frac{{\int_0^{{t_{\max }}} {{F_2}\left( {t,{t_0}} \right)\delta \left[ {x - {F_1}\left( {t,{t_0}} \right),y - {G_1}\left( {t,{t_0}} \right)} \right]dt} }}{{\int_0^{{t_{\max }}} {\delta \left[ {x - {F_1}\left( {t,{t_0}} \right),y - {G_1}\left( {t,{t_0}} \right)} \right]dt} }}\\ 
& = {F_2}\left( {t,{t_0}} \right){|_{t = F_1^{ - 1}\left( {x,{t_0}} \right)}},
\end{split}
\label{eq:C2}
\end{equation}
\end{small}
\begin{small}
\begin{equation}
\frac{{\partial \varphi \left( {x,y} \right)}}{{\partial y}} = {G_2}\left( {t,{t_0}} \right){|_{t = G_1^{ - 1}\left( {x,{t_0}} \right)}}.
\label{eq:C3}
\end{equation}
\end{small}

\noindent Since the relations of variables $x$, $y$ and variables $t$, ${t_0}$ are
\begin{small}
\begin{equation}
x = {F_1}\left( {t,{t_0}} \right) = f\left( {t + {t_0}} \right) - tf'\left( {t + {t_0}} \right),
\label{eq:C4}
\end{equation}
\end{small}
\begin{small}
\begin{equation}
y = {G_1}\left( {t,{t_0}} \right) = g\left( {t + {t_0}} \right) - tg'\left( {t + {t_0}} \right),
\label{eq:C5}
\end{equation}
\end{small}

\noindent the derivatives of phase distribution with respect to variables $t$ and ${t_0}$, can be obtained by applying the chain rule
\begin{small}
\begin{equation}
\begin{split}
\frac{{\partial \varphi \left( {x,y} \right)}}{{\partial t}} & = \frac{{\partial \varphi }}{{\partial x}} \cdot \frac{{\partial x}}{{\partial t}} + \frac{{\partial \varphi }}{{\partial y}} \cdot \frac{{\partial y}}{{\partial t}}\\ 
& =  - k{a^{ - 1}}t\left[ {f'\left( {t + {t_0}} \right)f''\left( {t + {t_0}} \right) }\right. \\
& \left. {~~~+ g'\left( {t + {t_0}} \right)g''\left( {t + {t_0}} \right)} \right],
\end{split}
\label{eq:C6}
\end{equation}
\end{small}
\begin{small}
\begin{equation}
\begin{split}
\frac{{\partial \varphi \left( {x,y} \right)}}{{\partial {t_0}}} & = \frac{{\partial \varphi }}{{\partial x}} \cdot \frac{{\partial x}}{{\partial {t_0}}} + \frac{{\partial \varphi }}{{\partial y}} \cdot \frac{{\partial y}}{{\partial {t_0}}}\\ 
& = k{a^{ - 1}}\left\{ {{{\left[ {f'\left( {t + {t_0}} \right)} \right]}^2} + {{\left[ {g'\left( {t + {t_0}} \right)} \right]}^2} }\right. \\
& \left.{~- t\left[ {f'\left( {t + {t_0}} \right)f''\left( {t + {t_0}} \right) + g'\left( {t + {t_0}} \right)g''\left( {t + {t_0}} \right)} \right]} \right\}.
\end{split}
\label{eq:C7}
\end{equation}
\end{small}

\noindent Furthermore, the mixed second derivatives can be figured out
\begin{small}
\begin{equation}
\begin{split}
\frac{{{\partial ^2}\varphi \left( {x,y} \right)}}{{\partial t\partial {t_0}}} = & - k{a^{ - 1}}t\left\{ {{{\left[ {f''\left( {t + {t_0}} \right)} \right]}^2} + f'\left( {t + {t_0}} \right)f'''\left( {t + {t_0}} \right) }\right. \\
& \left.{+ {{\left[ {g''\left( {t + {t_0}} \right)} \right]}^2} + g'\left( {t + {t_0}} \right)g'''\left( {t + {t_0}} \right)} \right\},
\end{split}
\label{eq:C8}
\end{equation}
\end{small}
\begin{small}
\begin{equation}
\begin{split}
\frac{{{\partial ^2}\varphi \left( {x,y} \right)}}{{\partial {t_0}\partial t}} = & - k{a^{ - 1}}t\left\{ {{{\left[ {f''\left( {t + {t_0}} \right)} \right]}^2} + f'\left( {t + {t_0}} \right)f'''\left( {t + {t_0}} \right) }\right. \\
& \left.{+ {{\left[ {g''\left( {t + {t_0}} \right)} \right]}^2} + g'\left( {t + {t_0}} \right)g'''\left( {t + {t_0}} \right)} \right\} \\
& + k{a^{ - 1}}\left[ {f'\left( {t + {t_0}} \right)f''\left( {t + {t_0}} \right) }\right. \\
& \left.{+ g'\left( {t + {t_0}} \right)g''\left( {t + {t_0}} \right)} \right].
\end{split}
\label{eq:C9}
\end{equation}
\end{small}

\noindent For a continuous function, these mixed second derivatives are equal, which leads to a constrain of
\begin{small}
\begin{equation}
f'\left( {t + {t_0}} \right)f''\left( {t + {t_0}} \right) + g'\left( {t + {t_0}} \right)g''\left( {t + {t_0}} \right) = 0,
\label{eq:C10}
\end{equation}
\end{small}

\noindent and can be further written as
\begin{small}
\begin{equation}
{\left[ {f'\left( {t + {t_0}} \right)} \right]^2} + {\left[ {g'\left( {t + {t_0}} \right)} \right]^2} = const \equiv {u^2}.
\label{eq:C11}
\end{equation}
\end{small}

\noindent This is the constraint presented in Eq.~(\ref{eq:10}). Moreover, bringing this condition back to Eqs.~(\ref{eq:C6}) and (\ref{eq:C7}) can derive the corresponding phase distribution
\begin{small}
\begin{equation}
\frac{{\partial \varphi \left( {x,y} \right)}}{{\partial t}} = 0,\frac{{\partial \varphi \left( {x,y} \right)}}{{\partial {t_0}}} = {u^2}k{a^{ - 1}} \Rightarrow \varphi \left( {x,y} \right) = {u^2}k{a^{ - 1}}{t_0}.
\label{eq:C12}
\end{equation}
\end{small}

Similarly, considering the Fourier space, we can obtain the angular spectrum including the intensity distribution
\begin{small}
\begin{equation}
\begin{split}
{{\rm P}^2}\left( {{k_x},{k_y}} \right) & = \iint{{W\left( {x,y,{k_x},{k_y}} \right)d{x}d{y}}} \\
& = \int_0^{{t_{\max }}} {\delta \left[ {{k_x} - {F_2}\left( {t,{t_0}} \right),{k_y} - {G_2}\left( {t,{t_0}} \right)} \right]dt},
\end{split} 
\label{eq:C13}
\end{equation}
\end{small}

\noindent and the derivatives of phase distribution with respect to spatial frequency ${k_x}$ and ${k_y}$
\begin{small}
\begin{equation}
\begin{split}
\frac{{\partial \Phi \left( {{k_x},{k_y}} \right)}}{{\partial {k_x}}} & = \frac{{\iint{{{x}W\left( {x,y,{k_x},{k_y}} \right)d{x}d{y}}}}}{{\iint{{W\left( {x,y,{k_x},{k_y}} \right)d{x}d{y}}}}} \\
&= {F_1}\left( {t,{t_0}} \right){|_{t = F_2^{ - 1}\left( {x,{t_0}} \right)}},
\end{split} 
\label{eq:C14}
\end{equation}
\end{small}
\begin{small}
\begin{equation}
\frac{{\partial \Phi \left( {{k_x},{k_y}} \right)}}{{\partial {k_y}}} = {G_1}\left( {t,{t_0}} \right){|_{t = G_2^{ - 1}\left( {x,{t_0}} \right)}}. 
\label{eq:C15}
\end{equation}
\end{small}

\noindent Since the relations of variables ${k_x}$, ${k_y}$ and variables $t$, $a$ are
\begin{small}
\begin{equation}
{k_x} = {F_2}\left( {t,{t_0}} \right) = k{a^{ - 1}}f'\left( {t + {t_0}} \right),
\label{eq:C16}
\end{equation}
\end{small}
\begin{small}
\begin{equation}
{k_y} = {G_2}\left( {t,{t_0}} \right) = k{a^{ - 1}}g'\left( {t + {t_0}} \right),
\label{eq:C17}
\end{equation}
\end{small}

\noindent the derivatives of phase distribution with respect to variables $t$ and $a$ can be obtained by applying the chain rule
\begin{small}
\begin{equation}
\begin{split}
\frac{{\partial \Phi \left( {{k_x},{k_y}} \right)}}{{\partial t}} & = \frac{{\partial \Phi }}{{\partial {k_x}}} \cdot \frac{{\partial {k_x}}}{{\partial t}} + \frac{{\partial \Phi }}{{\partial {k_y}}} \cdot \frac{{\partial {k_y}}}{{\partial t}} \\
& = - \left[ {f\left( {t + {t_0}} \right) - tf'\left( {t + {t_0}} \right)} \right] \cdot k{a^{ - 1}}f''\left( {t + {t_0}} \right) \\
& ~~~- \left[ {g\left( {t + {t_0}} \right) - tg'\left( {t + {t_0}} \right)} \right] \cdot k{a^{ - 1}}g''\left( {t + {t_0}} \right),
\end{split}
\label{eq:C18}
\end{equation}
\end{small}
\begin{small}
\begin{equation}
\begin{split}
\frac{{\partial \Phi \left( {{k_x},{k_y}} \right)}}{{\partial a}} & = \frac{{\partial \Phi }}{{\partial {k_x}}} \cdot \frac{{\partial {k_x}}}{{\partial a}} + \frac{{\partial \Phi }}{{\partial {k_y}}} \cdot \frac{{\partial {k_y}}}{{\partial a}} \\
& = \left[ {f\left( {t + {t_0}} \right) - tf'\left( {t + {t_0}} \right)} \right] \cdot k{a^{ - 2}}f'\left( {t + {t_0}} \right) \\
& ~~~+ \left[ {g\left( {t + {t_0}} \right) - tg'\left( {t + {t_0}} \right)} \right] \cdot k{a^{ - 2}}g'\left( {t + {t_0}} \right).
\end{split}
\label{eq:C19}
\end{equation}
\end{small}

\noindent Furthermore, the mixed second derivatives can be figured out
\begin{small}
\begin{equation}
\begin{split}
\frac{{{\partial ^2}\Phi \left( {{k_x},{k_y}} \right)}}{{\partial t\partial a}} = & \left[ {f\left( {t + {t_0}} \right) - tf'\left( {t + {t_0}} \right)} \right] \cdot k{a^{ - 2}}f''\left( {t + {t_0}} \right) \\
& + \left[ {g\left( {t + {t_0}} \right) - tg'\left( {t + {t_0}} \right)} \right] \cdot k{a^{ - 2}}g''\left( {t + {t_0}} \right),
\end{split}
\label{eq:C20}
\end{equation}
\end{small}
\begin{small}
\begin{equation}
\begin{split}
\frac{{{\partial ^2}\Phi \left( {{k_x},{k_y}} \right)}}{{\partial a\partial t}} = & \left[ {f\left( {t + {t_0}} \right) - tf'\left( {t + {t_0}} \right)} \right] \cdot k{a^{ - 2}}f''\left( {t + {t_0}} \right) \\
& + \left[ {g\left( {t + {t_0}} \right) - tg'\left( {t + {t_0}} \right)} \right] \cdot k{a^{ - 2}}g''\left( {t + {t_0}} \right) \\
& - k{a^{ - 2}}t\left[ {f'\left( {t + {t_0}} \right)f''\left( {t + {t_0}} \right) }\right. \\
& \left.{+ g'\left( {t + {t_0}} \right)g''\left( {t + {t_0}} \right)} \right].
\end{split}
\label{eq:C21}
\end{equation}
\end{small}

\noindent The equality of these mixed second derivatives also lead to the same constrain of Eq.~(\ref{eq:C11}). Bringing this condition back to Eqs.~(\ref{eq:C18}) and (\ref{eq:C19}) can also figure out the corresponding phase distribution
\begin{small}
\begin{equation}
\begin{split}
\Phi \left( {{k_x},{k_y}} \right) = & k{a^{ - 1}}\left[ {{u^2}t - f\left( {t + {t_0}} \right)f'\left( {t + {t_0}} \right) }\right. \\
& \left.{ - g\left( {t + {t_0}} \right)g'\left( {t + {t_0}} \right)} \right].
\end{split}
\label{eq:C22}
\end{equation}
\end{small}


\begin{thebibliography}{37}%
\makeatletter
\providecommand \@ifxundefined [1]{%
 \@ifx{#1\undefined}
}%
\providecommand \@ifnum [1]{%
 \ifnum #1\expandafter \@firstoftwo
 \else \expandafter \@secondoftwo
 \fi
}%
\providecommand \@ifx [1]{%
 \ifx #1\expandafter \@firstoftwo
 \else \expandafter \@secondoftwo
 \fi
}%
\providecommand \natexlab [1]{#1}%
\providecommand \enquote  [1]{``#1''}%
\providecommand \bibnamefont  [1]{#1}%
\providecommand \bibfnamefont [1]{#1}%
\providecommand \citenamefont [1]{#1}%
\providecommand \href@noop [0]{\@secondoftwo}%
\providecommand \href [0]{\begingroup \@sanitize@url \@href}%
\providecommand \@href[1]{\@@startlink{#1}\@@href}%
\providecommand \@@href[1]{\endgroup#1\@@endlink}%
\providecommand \@sanitize@url [0]{\catcode `\\12\catcode `\$12\catcode
  `\&12\catcode `\#12\catcode `\^12\catcode `\_12\catcode `\%12\relax}%
\providecommand \@@startlink[1]{}%
\providecommand \@@endlink[0]{}%
\providecommand \url  [0]{\begingroup\@sanitize@url \@url }%
\providecommand \@url [1]{\endgroup\@href {#1}{\urlprefix }}%
\providecommand \urlprefix  [0]{URL }%
\providecommand \Eprint [0]{\href }%
\@ifxundefined \urlstyle {%
  \providecommand \doi  [0]{\begingroup \@sanitize@url \@doi}%
  \providecommand \@doi [1]{\endgroup \@@startlink {\doibase
  #1}doi:\discretionary {}{}{}#1\@@endlink }%
}{%
  \providecommand \doi  [0]{doi:\discretionary{}{}{}\begingroup
  \urlstyle{rm}\Url }%
}%
\providecommand \doibase [0]{http://dx.doi.org/}%
\providecommand \Doi [0]{\begingroup \@sanitize@url \@Doi }%
\providecommand \@Doi  [1]{\endgroup\@@startlink{\doibase#1}\@@Doi}%
\providecommand \@@Doi [1]{#1\@@endlink}%
\providecommand \selectlanguage [0]{\@gobble}%
\providecommand \bibinfo  [0]{\@secondoftwo}%
\providecommand \bibfield  [0]{\@secondoftwo}%
\providecommand \translation [1]{[#1]}%
\providecommand \BibitemOpen [0]{}%
\providecommand \bibitemStop [0]{}%
\providecommand \bibitemNoStop [0]{.\EOS\space}%
\providecommand \EOS [0]{\spacefactor3000\relax}%
\providecommand \BibitemShut  [1]{\csname bibitem#1\endcsname}%
\bibitem [{\citenamefont {Gansel}\ \emph {et~al.}(2009)\citenamefont {Gansel},
  \citenamefont {Thiel}, \citenamefont {Rill}, \citenamefont {Decker},
  \citenamefont {Bade}, \citenamefont {Saile}, \citenamefont {von Freymann},
  \citenamefont {Linden},\ and\ \citenamefont {Wegener}}]{Gansel1513}%
  \BibitemOpen
  \bibfield  {author} {\bibinfo {author} {\bibfnamefont {J.~K.}\ \bibnamefont
  {Gansel}}, \bibinfo {author} {\bibfnamefont {M.}~\bibnamefont {Thiel}},
  \bibinfo {author} {\bibfnamefont {M.~S.}\ \bibnamefont {Rill}}, \bibinfo
  {author} {\bibfnamefont {M.}~\bibnamefont {Decker}}, \bibinfo {author}
  {\bibfnamefont {K.}~\bibnamefont {Bade}}, \bibinfo {author} {\bibfnamefont
  {V.}~\bibnamefont {Saile}}, \bibinfo {author} {\bibfnamefont
  {G.}~\bibnamefont {von Freymann}}, \bibinfo {author} {\bibfnamefont
  {S.}~\bibnamefont {Linden}}, \ and\ \bibinfo {author} {\bibfnamefont
  {M.}~\bibnamefont {Wegener}},\ }\Doi {10.1126/science.1177031} {\bibfield
  {journal} {\bibinfo  {journal} {Science},\ }\textbf {\bibinfo {volume}
  {325}},\ \bibinfo {pages} {1513} (\bibinfo {year} {2009})}\BibitemShut
  {NoStop}%
\bibitem [{\citenamefont {Yu}\ \emph {et~al.}(2011)\citenamefont {Yu},
  \citenamefont {Genevet}, \citenamefont {Kats}, \citenamefont {Aieta},
  \citenamefont {Tetienne}, \citenamefont {Capasso},\ and\ \citenamefont
  {Gaburro}}]{Yu333}%
  \BibitemOpen
  \bibfield  {author} {\bibinfo {author} {\bibfnamefont {N.}~\bibnamefont
  {Yu}}, \bibinfo {author} {\bibfnamefont {P.}~\bibnamefont {Genevet}},
  \bibinfo {author} {\bibfnamefont {M.~A.}\ \bibnamefont {Kats}}, \bibinfo
  {author} {\bibfnamefont {F.}~\bibnamefont {Aieta}}, \bibinfo {author}
  {\bibfnamefont {J.-P.}\ \bibnamefont {Tetienne}}, \bibinfo {author}
  {\bibfnamefont {F.}~\bibnamefont {Capasso}}, \ and\ \bibinfo {author}
  {\bibfnamefont {Z.}~\bibnamefont {Gaburro}},\ }\Doi {10.1126/science.1210713}
  {\textbf {\bibinfo {volume} {334}},\ \bibinfo {pages} {333} (\bibinfo {year}
  {2011})}\BibitemShut {NoStop}%
\bibitem [{\citenamefont {Lin}\ \emph {et~al.}(2014)\citenamefont {Lin},
  \citenamefont {Fan}, \citenamefont {Hasman},\ and\ \citenamefont
  {Brongersma}}]{Lin298}%
  \BibitemOpen
  \bibfield  {author} {\bibinfo {author} {\bibfnamefont {D.}~\bibnamefont
  {Lin}}, \bibinfo {author} {\bibfnamefont {P.}~\bibnamefont {Fan}}, \bibinfo
  {author} {\bibfnamefont {E.}~\bibnamefont {Hasman}}, \ and\ \bibinfo {author}
  {\bibfnamefont {M.~L.}\ \bibnamefont {Brongersma}},\ }\Doi
  {10.1126/science.1253213} {\bibfield  {journal} {\bibinfo  {journal}
  {Science},\ }\textbf {\bibinfo {volume} {345}},\ \bibinfo {pages} {298}
  (\bibinfo {year} {2014})}\BibitemShut {NoStop}%
\bibitem [{\citenamefont {Dolev}\ \emph {et~al.}(2012)\citenamefont {Dolev},
  \citenamefont {Epstein},\ and\ \citenamefont
  {Arie}}]{PhysRevLett.109.203903}%
  \BibitemOpen
  \bibfield  {author} {\bibinfo {author} {\bibfnamefont {I.}~\bibnamefont
  {Dolev}}, \bibinfo {author} {\bibfnamefont {I.}~\bibnamefont {Epstein}}, \
  and\ \bibinfo {author} {\bibfnamefont {A.}~\bibnamefont {Arie}},\ }\Doi
  {10.1103/PhysRevLett.109.203903} {\bibfield  {journal} {\bibinfo  {journal}
  {Phys. Rev. Lett.},\ }\textbf {\bibinfo {volume} {109}},\ \bibinfo {pages}
  {203903} (\bibinfo {year} {2012})}\BibitemShut {NoStop}%
\bibitem [{\citenamefont {Maguid}\ \emph {et~al.}(2016)\citenamefont {Maguid},
  \citenamefont {Yulevich}, \citenamefont {Veksler}, \citenamefont {Kleiner},
  \citenamefont {Brongersma},\ and\ \citenamefont {Hasman}}]{Maguid1202}%
  \BibitemOpen
  \bibfield  {author} {\bibinfo {author} {\bibfnamefont {E.}~\bibnamefont
  {Maguid}}, \bibinfo {author} {\bibfnamefont {I.}~\bibnamefont {Yulevich}},
  \bibinfo {author} {\bibfnamefont {D.}~\bibnamefont {Veksler}}, \bibinfo
  {author} {\bibfnamefont {V.}~\bibnamefont {Kleiner}}, \bibinfo {author}
  {\bibfnamefont {M.~L.}\ \bibnamefont {Brongersma}}, \ and\ \bibinfo {author}
  {\bibfnamefont {E.}~\bibnamefont {Hasman}},\ }\Doi {10.1126/science.aaf3417}
  {\bibfield  {journal} {\bibinfo  {journal} {Science},\ }\textbf {\bibinfo
  {volume} {352}},\ \bibinfo {pages} {1202} (\bibinfo {year}
  {2016})}\BibitemShut {NoStop}%
\bibitem [{\citenamefont {Yu}\ and\ \citenamefont {Capasso}(2014)}]{Nat.Mater}%
  \BibitemOpen
  \bibfield  {author} {\bibinfo {author} {\bibfnamefont {N.}~\bibnamefont
  {Yu}}\ and\ \bibinfo {author} {\bibfnamefont {F.}~\bibnamefont {Capasso}},\
  }\Doi {10.1038/nmat3839} {\bibfield  {journal} {\bibinfo  {journal} {Nat.
  Mater.},\ }\textbf {\bibinfo {volume} {13}},\ \bibinfo {pages} {139}
  (\bibinfo {year} {2014})}\BibitemShut {NoStop}%
\bibitem [{\citenamefont {Siviloglou}\ and\ \citenamefont
  {Christodoulides}(2007)}]{Siviloglou:07}%
  \BibitemOpen
  \bibfield  {author} {\bibinfo {author} {\bibfnamefont {G.~A.}\ \bibnamefont
  {Siviloglou}}\ and\ \bibinfo {author} {\bibfnamefont {D.~N.}\ \bibnamefont
  {Christodoulides}},\ }\Doi {10.1364/OL.32.000979} {\bibfield  {journal}
  {\bibinfo  {journal} {Opt. Lett.},\ }\textbf {\bibinfo {volume} {32}},\
  \bibinfo {pages} {979} (\bibinfo {year} {2007})}\BibitemShut {NoStop}%
\bibitem [{\citenamefont {Siviloglou}\ \emph {et~al.}(2007)\citenamefont
  {Siviloglou}, \citenamefont {Broky}, \citenamefont {Dogariu},\ and\
  \citenamefont {Christodoulides}}]{PhysRevLett.99.213901}%
  \BibitemOpen
  \bibfield  {author} {\bibinfo {author} {\bibfnamefont {G.~A.}\ \bibnamefont
  {Siviloglou}}, \bibinfo {author} {\bibfnamefont {J.}~\bibnamefont {Broky}},
  \bibinfo {author} {\bibfnamefont {A.}~\bibnamefont {Dogariu}}, \ and\
  \bibinfo {author} {\bibfnamefont {D.~N.}\ \bibnamefont {Christodoulides}},\
  }\Doi {10.1103/PhysRevLett.99.213901} {\bibfield  {journal} {\bibinfo
  {journal} {Phys. Rev. Lett.},\ }\textbf {\bibinfo {volume} {99}},\ \bibinfo
  {pages} {213901} (\bibinfo {year} {2007})}\BibitemShut {NoStop}%
\bibitem [{\citenamefont {Baumgartl}\ \emph {et~al.}(2008)\citenamefont
  {Baumgartl}, \citenamefont {Mazilu},\ and\ \citenamefont
  {Dholakia}}]{Nat.Photon.2008.201}%
  \BibitemOpen
  \bibfield  {author} {\bibinfo {author} {\bibfnamefont {J.}~\bibnamefont
  {Baumgartl}}, \bibinfo {author} {\bibfnamefont {M.}~\bibnamefont {Mazilu}}, \
  and\ \bibinfo {author} {\bibfnamefont {K.}~\bibnamefont {Dholakia}},\ }\Doi
  {10.1038/nphoton.2008.201} {\bibfield  {journal} {\bibinfo  {journal} {Nat.
  Photon.},\ }\textbf {\bibinfo {volume} {2}},\ \bibinfo {pages} {675}
  (\bibinfo {year} {2008})}\BibitemShut {NoStop}%
\bibitem [{\citenamefont {Zhao}\ \emph {et~al.}(2015)\citenamefont {Zhao},
  \citenamefont {Chremmos}, \citenamefont {Song}, \citenamefont
  {Christodoulides}, \citenamefont {Efremidis},\ and\ \citenamefont
  {Chen}}]{Zhao2015Curved}%
  \BibitemOpen
  \bibfield  {author} {\bibinfo {author} {\bibfnamefont {J.}~\bibnamefont
  {Zhao}}, \bibinfo {author} {\bibfnamefont {I.~D.}\ \bibnamefont {Chremmos}},
  \bibinfo {author} {\bibfnamefont {D.}~\bibnamefont {Song}}, \bibinfo {author}
  {\bibfnamefont {D.~N.}\ \bibnamefont {Christodoulides}}, \bibinfo {author}
  {\bibfnamefont {N.~K.}\ \bibnamefont {Efremidis}}, \ and\ \bibinfo {author}
  {\bibfnamefont {Z.}~\bibnamefont {Chen}},\ }\Doi {10.1038/srep12086}
  {\bibfield  {journal} {\bibinfo  {journal} {Sci. Rep.},\ }\textbf {\bibinfo
  {volume} {5}},\ \bibinfo {pages} {12086} (\bibinfo {year}
  {2015})}\BibitemShut {NoStop}%
\bibitem [{\citenamefont {Mathis}\ \emph {et~al.}(2012)\citenamefont {Mathis},
  \citenamefont {Courvoisier}, \citenamefont {Froehly}, \citenamefont
  {Furfaro}, \citenamefont {Jacquot}, \citenamefont {Lacourt},\ and\
  \citenamefont {Dudley}}]{apl}%
  \BibitemOpen
  \bibfield  {author} {\bibinfo {author} {\bibfnamefont {A.}~\bibnamefont
  {Mathis}}, \bibinfo {author} {\bibfnamefont {F.}~\bibnamefont {Courvoisier}},
  \bibinfo {author} {\bibfnamefont {L.}~\bibnamefont {Froehly}}, \bibinfo
  {author} {\bibfnamefont {L.}~\bibnamefont {Furfaro}}, \bibinfo {author}
  {\bibfnamefont {M.}~\bibnamefont {Jacquot}}, \bibinfo {author} {\bibfnamefont
  {P.~A.}\ \bibnamefont {Lacourt}}, \ and\ \bibinfo {author} {\bibfnamefont
  {J.~M.}\ \bibnamefont {Dudley}},\ }\href
  {http://scitation.aip.org/content/aip/journal/apl/101/7/10.1063/1.4745925}
  {\bibfield  {journal} {\bibinfo  {journal} {Appl. Phys. Lett.},\ }\textbf
  {\bibinfo {volume} {101}},\ \bibinfo {pages} {071110} (\bibinfo {year}
  {2012})}\BibitemShut {NoStop}%
\bibitem [{\citenamefont {Jia}\ \emph {et~al.}(2014)\citenamefont {Jia},
  \citenamefont {Vaughan},\ and\ \citenamefont {Zhuang}}]{Nat.PhotonJS}%
  \BibitemOpen
  \bibfield  {author} {\bibinfo {author} {\bibfnamefont {S.}~\bibnamefont
  {Jia}}, \bibinfo {author} {\bibfnamefont {J.~C.}\ \bibnamefont {Vaughan}}, \
  and\ \bibinfo {author} {\bibfnamefont {X.}~\bibnamefont {Zhuang}},\ }\Doi
  {10.1038/nphoton.2014.13} {\bibfield  {journal} {\bibinfo  {journal} {Nat.
  Photon.},\ }\textbf {\bibinfo {volume} {8}},\ \bibinfo {pages} {302}
  (\bibinfo {year} {2014})}\BibitemShut {NoStop}%
\bibitem [{\citenamefont {Vettenburg}\ \emph {et~al.}(2014)\citenamefont
  {Vettenburg}, \citenamefont {Dalgarno}, \citenamefont {Nylk}, \citenamefont
  {Coll-Llad\'{o}}, \citenamefont {Ferrier}, \citenamefont
  {\v{C}i\v{z}m\'{a}r}, \citenamefont {Gunn-Moore},\ and\ \citenamefont
  {Dholakia}}]{Nat.Meth}%
  \BibitemOpen
  \bibfield  {author} {\bibinfo {author} {\bibfnamefont {T.}~\bibnamefont
  {Vettenburg}}, \bibinfo {author} {\bibfnamefont {H.~I.~C.}\ \bibnamefont
  {Dalgarno}}, \bibinfo {author} {\bibfnamefont {J.}~\bibnamefont {Nylk}},
  \bibinfo {author} {\bibfnamefont {C.}~\bibnamefont {Coll-Llad\'{o}}},
  \bibinfo {author} {\bibfnamefont {D.~E.~K.}\ \bibnamefont {Ferrier}},
  \bibinfo {author} {\bibfnamefont {T.}~\bibnamefont {\v{C}i\v{z}m\'{a}r}},
  \bibinfo {author} {\bibfnamefont {F.~J.}\ \bibnamefont {Gunn-Moore}}, \ and\
  \bibinfo {author} {\bibfnamefont {K.}~\bibnamefont {Dholakia}},\ }\Doi
  {10.1038/nmeth.2922} {\bibfield  {journal} {\bibinfo  {journal} {Nat.
  Method},\ }\textbf {\bibinfo {volume} {11}},\ \bibinfo {pages} {541}
  (\bibinfo {year} {2014})}\BibitemShut {NoStop}%
\bibitem [{\citenamefont {Rose}\ \emph {et~al.}(2013)\citenamefont {Rose},
  \citenamefont {Diebel}, \citenamefont {Boguslawski},\ and\ \citenamefont
  {Denz}}]{aplR}%
  \BibitemOpen
  \bibfield  {author} {\bibinfo {author} {\bibfnamefont {P.}~\bibnamefont
  {Rose}}, \bibinfo {author} {\bibfnamefont {F.}~\bibnamefont {Diebel}},
  \bibinfo {author} {\bibfnamefont {M.}~\bibnamefont {Boguslawski}}, \ and\
  \bibinfo {author} {\bibfnamefont {C.}~\bibnamefont {Denz}},\ }\href
  {http://scitation.aip.org/content/aip/journal/apl/102/10/10.1063/1.4793668}
  {\bibfield  {journal} {\bibinfo  {journal} {Appl. Phys. Lett.},\ }\textbf
  {\bibinfo {volume} {102}},\ \bibinfo {pages} {101101} (\bibinfo {year}
  {2013})}\BibitemShut {NoStop}%
\bibitem [{\citenamefont {Chong}\ \emph {et~al.}(2010)\citenamefont {Chong},
  \citenamefont {Renninger}, \citenamefont {Christodoulides},\ and\
  \citenamefont {Wise}}]{chong2010airy}%
  \BibitemOpen
  \bibfield  {author} {\bibinfo {author} {\bibfnamefont {A.}~\bibnamefont
  {Chong}}, \bibinfo {author} {\bibfnamefont {W.~H.}\ \bibnamefont
  {Renninger}}, \bibinfo {author} {\bibfnamefont {D.~N.}\ \bibnamefont
  {Christodoulides}}, \ and\ \bibinfo {author} {\bibfnamefont {F.~W.}\
  \bibnamefont {Wise}},\ }\Doi {10.1038/nphoton.2009.264} {\bibfield  {journal}
  {\bibinfo  {journal} {Nat. Photon.},\ }\textbf {\bibinfo {volume} {4}},\
  \bibinfo {pages} {103} (\bibinfo {year} {2010})}\BibitemShut {NoStop}%
\bibitem [{\citenamefont {Abdollahpour}\ \emph {et~al.}(2010)\citenamefont
  {Abdollahpour}, \citenamefont {Suntsov}, \citenamefont {Papazoglou},\ and\
  \citenamefont {Tzortzakis}}]{PhysRevLett.105.253901}%
  \BibitemOpen
  \bibfield  {author} {\bibinfo {author} {\bibfnamefont {D.}~\bibnamefont
  {Abdollahpour}}, \bibinfo {author} {\bibfnamefont {S.}~\bibnamefont
  {Suntsov}}, \bibinfo {author} {\bibfnamefont {D.~G.}\ \bibnamefont
  {Papazoglou}}, \ and\ \bibinfo {author} {\bibfnamefont {S.}~\bibnamefont
  {Tzortzakis}},\ }\Doi {10.1103/PhysRevLett.105.253901} {\bibfield  {journal}
  {\bibinfo  {journal} {Phys. Rev. Lett.},\ }\textbf {\bibinfo {volume}
  {105}},\ \bibinfo {pages} {253901} (\bibinfo {year} {2010})}\BibitemShut
  {NoStop}%
\bibitem [{\citenamefont {Clerici}\ \emph {et~al.}(2015)\citenamefont
  {Clerici}, \citenamefont {Hu}, \citenamefont {Lassonde}, \citenamefont
  {Mili{\'a}n}, \citenamefont {Couairon}, \citenamefont {Christodoulides},
  \citenamefont {Chen}, \citenamefont {Razzari}, \citenamefont {Vidal},
  \citenamefont {L{\'e}gar{\'e}}, \citenamefont {Faccio},\ and\ \citenamefont
  {Morandotti}}]{Clericie1400111}%
  \BibitemOpen
  \bibfield  {author} {\bibinfo {author} {\bibfnamefont {M.}~\bibnamefont
  {Clerici}}, \bibinfo {author} {\bibfnamefont {Y.}~\bibnamefont {Hu}},
  \bibinfo {author} {\bibfnamefont {P.}~\bibnamefont {Lassonde}}, \bibinfo
  {author} {\bibfnamefont {C.}~\bibnamefont {Mili{\'a}n}}, \bibinfo {author}
  {\bibfnamefont {A.}~\bibnamefont {Couairon}}, \bibinfo {author}
  {\bibfnamefont {D.~N.}\ \bibnamefont {Christodoulides}}, \bibinfo {author}
  {\bibfnamefont {Z.}~\bibnamefont {Chen}}, \bibinfo {author} {\bibfnamefont
  {L.}~\bibnamefont {Razzari}}, \bibinfo {author} {\bibfnamefont
  {F.}~\bibnamefont {Vidal}}, \bibinfo {author} {\bibfnamefont
  {F.}~\bibnamefont {L{\'e}gar{\'e}}}, \bibinfo {author} {\bibfnamefont
  {D.}~\bibnamefont {Faccio}}, \ and\ \bibinfo {author} {\bibfnamefont
  {R.}~\bibnamefont {Morandotti}},\ }\href
  {http://advances.sciencemag.org/content/1/5/e1400111} {\bibfield  {journal}
  {\bibinfo  {journal} {Sci. Adv.},\ }\textbf {\bibinfo {volume} {1}},\
  \bibinfo {pages} {1400111} (\bibinfo {year} {2015})}\BibitemShut {NoStop}%
\bibitem [{\citenamefont {Polynkin}\ \emph {et~al.}(2009)\citenamefont
  {Polynkin}, \citenamefont {Kolesik}, \citenamefont {Moloney}, \citenamefont
  {Siviloglou},\ and\ \citenamefont {Christodoulides}}]{Polynkin229}%
  \BibitemOpen
  \bibfield  {author} {\bibinfo {author} {\bibfnamefont {P.}~\bibnamefont
  {Polynkin}}, \bibinfo {author} {\bibfnamefont {M.}~\bibnamefont {Kolesik}},
  \bibinfo {author} {\bibfnamefont {J.~V.}\ \bibnamefont {Moloney}}, \bibinfo
  {author} {\bibfnamefont {G.~A.}\ \bibnamefont {Siviloglou}}, \ and\ \bibinfo
  {author} {\bibfnamefont {D.~N.}\ \bibnamefont {Christodoulides}},\ }\Doi
  {10.1126/science.1169544} {\bibfield  {journal} {\bibinfo  {journal}
  {Science},\ }\textbf {\bibinfo {volume} {324}},\ \bibinfo {pages} {229}
  (\bibinfo {year} {2009})}\BibitemShut {NoStop}%
\bibitem [{\citenamefont {Zhang}\ \emph {et~al.}(2014)\citenamefont {Zhang},
  \citenamefont {Li}, \citenamefont {Zhu}, \citenamefont {Zhu}, \citenamefont
  {Yang}, \citenamefont {Wang}, \citenamefont {Yin},\ and\ \citenamefont
  {Zhang}}]{Nat.Common}%
  \BibitemOpen
  \bibfield  {author} {\bibinfo {author} {\bibfnamefont {P.}~\bibnamefont
  {Zhang}}, \bibinfo {author} {\bibfnamefont {T.}~\bibnamefont {Li}}, \bibinfo
  {author} {\bibfnamefont {J.}~\bibnamefont {Zhu}}, \bibinfo {author}
  {\bibfnamefont {X.}~\bibnamefont {Zhu}}, \bibinfo {author} {\bibfnamefont
  {S.}~\bibnamefont {Yang}}, \bibinfo {author} {\bibfnamefont {Y.}~\bibnamefont
  {Wang}}, \bibinfo {author} {\bibfnamefont {X.}~\bibnamefont {Yin}}, \ and\
  \bibinfo {author} {\bibfnamefont {X.}~\bibnamefont {Zhang}},\ }\Doi
  {10.1038/ncomms5316} {\bibfield  {journal} {\bibinfo  {journal} {Nat.
  Commun.},\ }\textbf {\bibinfo {volume} {5}},\ \bibinfo {pages} {4316}
  (\bibinfo {year} {2014})}\BibitemShut {NoStop}%
\bibitem [{\citenamefont {Voloch-Bloch}\ \emph {et~al.}(2013)\citenamefont
  {Voloch-Bloch}, \citenamefont {Lereah}, \citenamefont {Lilach}, \citenamefont
  {Gover},\ and\ \citenamefont {Arie}}]{Nature}%
  \BibitemOpen
  \bibfield  {author} {\bibinfo {author} {\bibfnamefont {N.}~\bibnamefont
  {Voloch-Bloch}}, \bibinfo {author} {\bibfnamefont {Y.}~\bibnamefont
  {Lereah}}, \bibinfo {author} {\bibfnamefont {Y.}~\bibnamefont {Lilach}},
  \bibinfo {author} {\bibfnamefont {A.}~\bibnamefont {Gover}}, \ and\ \bibinfo
  {author} {\bibfnamefont {A.}~\bibnamefont {Arie}},\ }\Doi
  {10.1038/nature11840} {\bibfield  {journal} {\bibinfo  {journal} {Nature},\
  }\textbf {\bibinfo {volume} {494}},\ \bibinfo {pages} {7437} (\bibinfo {year}
  {2013})}\BibitemShut {NoStop}%
\bibitem [{\citenamefont {Minovich}\ \emph {et~al.}(2011)\citenamefont
  {Minovich}, \citenamefont {Klein}, \citenamefont {Janunts}, \citenamefont
  {Pertsch}, \citenamefont {Neshev},\ and\ \citenamefont
  {Kivshar}}]{PhysRevLett.107.116802}%
  \BibitemOpen
  \bibfield  {author} {\bibinfo {author} {\bibfnamefont {A.}~\bibnamefont
  {Minovich}}, \bibinfo {author} {\bibfnamefont {A.~E.}\ \bibnamefont {Klein}},
  \bibinfo {author} {\bibfnamefont {N.}~\bibnamefont {Janunts}}, \bibinfo
  {author} {\bibfnamefont {T.}~\bibnamefont {Pertsch}}, \bibinfo {author}
  {\bibfnamefont {D.~N.}\ \bibnamefont {Neshev}}, \ and\ \bibinfo {author}
  {\bibfnamefont {Y.~S.}\ \bibnamefont {Kivshar}},\ }\Doi
  {10.1103/PhysRevLett.107.116802} {\bibfield  {journal} {\bibinfo  {journal}
  {Phys. Rev. Lett.},\ }\textbf {\bibinfo {volume} {107}},\ \bibinfo {pages}
  {116802} (\bibinfo {year} {2011})}\BibitemShut {NoStop}%
\bibitem [{\citenamefont {Li}\ \emph {et~al.}(2011)\citenamefont {Li},
  \citenamefont {Li}, \citenamefont {Wang}, \citenamefont {Zhang},\ and\
  \citenamefont {Zhu}}]{PhysRevLett.107.126804}%
  \BibitemOpen
  \bibfield  {author} {\bibinfo {author} {\bibfnamefont {L.}~\bibnamefont
  {Li}}, \bibinfo {author} {\bibfnamefont {T.}~\bibnamefont {Li}}, \bibinfo
  {author} {\bibfnamefont {S.~M.}\ \bibnamefont {Wang}}, \bibinfo {author}
  {\bibfnamefont {C.}~\bibnamefont {Zhang}}, \ and\ \bibinfo {author}
  {\bibfnamefont {S.~N.}\ \bibnamefont {Zhu}},\ }\Doi
  {10.1103/PhysRevLett.107.126804} {\bibfield  {journal} {\bibinfo  {journal}
  {Phys. Rev. Lett.},\ }\textbf {\bibinfo {volume} {107}},\ \bibinfo {pages}
  {126804} (\bibinfo {year} {2011})}\BibitemShut {NoStop}%
\bibitem [{\citenamefont {Epstein}\ and\ \citenamefont
  {Arie}(2014)}]{PhysRevLett.112.023903}%
  \BibitemOpen
  \bibfield  {author} {\bibinfo {author} {\bibfnamefont {I.}~\bibnamefont
  {Epstein}}\ and\ \bibinfo {author} {\bibfnamefont {A.}~\bibnamefont {Arie}},\
  }\Doi {10.1103/PhysRevLett.112.023903} {\bibfield  {journal} {\bibinfo
  {journal} {Phys. Rev. Lett.},\ }\textbf {\bibinfo {volume} {112}},\ \bibinfo
  {pages} {023903} (\bibinfo {year} {2014})}\BibitemShut {NoStop}%
\bibitem [{\citenamefont {Berry}\ and\ \citenamefont
  {Upstill}(1980)}]{Berry1980257}%
  \BibitemOpen
  \bibfield  {author} {\bibinfo {author} {\bibfnamefont {M.}~\bibnamefont
  {Berry}}\ and\ \bibinfo {author} {\bibfnamefont {C.}~\bibnamefont {Upstill}}\
  }(\bibinfo  {publisher} {Elsevier},\ \bibinfo {year} {1980})\ pp.\ \bibinfo
  {pages} {257--346}\BibitemShut {NoStop}%
\bibitem [{\citenamefont {Greenfield}\ \emph {et~al.}(2011)\citenamefont
  {Greenfield}, \citenamefont {Segev}, \citenamefont {Walasik},\ and\
  \citenamefont {Raz}}]{PhysRevLett.106.213902}%
  \BibitemOpen
  \bibfield  {author} {\bibinfo {author} {\bibfnamefont {E.}~\bibnamefont
  {Greenfield}}, \bibinfo {author} {\bibfnamefont {M.}~\bibnamefont {Segev}},
  \bibinfo {author} {\bibfnamefont {W.}~\bibnamefont {Walasik}}, \ and\
  \bibinfo {author} {\bibfnamefont {O.}~\bibnamefont {Raz}},\ }\Doi
  {10.1103/PhysRevLett.106.213902} {\bibfield  {journal} {\bibinfo  {journal}
  {Phys. Rev. Lett.},\ }\textbf {\bibinfo {volume} {106}},\ \bibinfo {pages}
  {213902} (\bibinfo {year} {2011})}\BibitemShut {NoStop}%
\bibitem [{\citenamefont {Froehly}\ \emph {et~al.}(2011)\citenamefont
  {Froehly}, \citenamefont {Courvoisier}, \citenamefont {Mathis}, \citenamefont
  {Jacquot}, \citenamefont {Furfaro}, \citenamefont {Giust}, \citenamefont
  {Lacourt},\ and\ \citenamefont {Dudley}}]{Froehly:11}%
  \BibitemOpen
  \bibfield  {author} {\bibinfo {author} {\bibfnamefont {L.}~\bibnamefont
  {Froehly}}, \bibinfo {author} {\bibfnamefont {F.}~\bibnamefont
  {Courvoisier}}, \bibinfo {author} {\bibfnamefont {A.}~\bibnamefont {Mathis}},
  \bibinfo {author} {\bibfnamefont {M.}~\bibnamefont {Jacquot}}, \bibinfo
  {author} {\bibfnamefont {L.}~\bibnamefont {Furfaro}}, \bibinfo {author}
  {\bibfnamefont {R.}~\bibnamefont {Giust}}, \bibinfo {author} {\bibfnamefont
  {P.~A.}\ \bibnamefont {Lacourt}}, \ and\ \bibinfo {author} {\bibfnamefont
  {J.~M.}\ \bibnamefont {Dudley}},\ }\Doi {10.1364/OE.19.016455} {\bibfield
  {journal} {\bibinfo  {journal} {Opt. Express},\ }\textbf {\bibinfo {volume}
  {19}},\ \bibinfo {pages} {16455} (\bibinfo {year} {2011})}\BibitemShut
  {NoStop}%
\bibitem [{\citenamefont {Hu}\ \emph {et~al.}(2013)\citenamefont {Hu},
  \citenamefont {Bongiovanni}, \citenamefont {Chen},\ and\ \citenamefont
  {Morandotti}}]{PhysRevA.88.043809}%
  \BibitemOpen
  \bibfield  {author} {\bibinfo {author} {\bibfnamefont {Y.}~\bibnamefont
  {Hu}}, \bibinfo {author} {\bibfnamefont {D.}~\bibnamefont {Bongiovanni}},
  \bibinfo {author} {\bibfnamefont {Z.}~\bibnamefont {Chen}}, \ and\ \bibinfo
  {author} {\bibfnamefont {R.}~\bibnamefont {Morandotti}},\ }\Doi
  {10.1103/PhysRevA.88.043809} {\bibfield  {journal} {\bibinfo  {journal}
  {Phys. Rev. A},\ }\textbf {\bibinfo {volume} {88}},\ \bibinfo {pages}
  {043809} (\bibinfo {year} {2013})}\BibitemShut {NoStop}%
\bibitem [{\citenamefont {Bongiovanni}\ \emph {et~al.}(2015)\citenamefont
  {Bongiovanni}, \citenamefont {Hu}, \citenamefont {Wetzel}, \citenamefont
  {Robles}, \citenamefont {Mendoza}, \citenamefont {Martipaname帽o},
  \citenamefont {Chen},\ and\ \citenamefont
  {Morandotti}}]{Bongiovanni2015Efficient}%
  \BibitemOpen
  \bibfield  {author} {\bibinfo {author} {\bibfnamefont {D.}~\bibnamefont
  {Bongiovanni}}, \bibinfo {author} {\bibfnamefont {Y.}~\bibnamefont {Hu}},
  \bibinfo {author} {\bibfnamefont {B.}~\bibnamefont {Wetzel}}, \bibinfo
  {author} {\bibfnamefont {R.~A.}\ \bibnamefont {Robles}}, \bibinfo {author}
  {\bibfnamefont {G.~G.}\ \bibnamefont {Mendoza}}, \bibinfo {author}
  {\bibfnamefont {E.~A.}\ \bibnamefont {Martipaname帽o}}, \bibinfo {author}
  {\bibfnamefont {Z.}~\bibnamefont {Chen}}, \ and\ \bibinfo {author}
  {\bibfnamefont {R.}~\bibnamefont {Morandotti}},\ }\Doi {10.1038/srep13197}
  {\bibfield  {journal} {\bibinfo  {journal} {Sci. Rep.},\ }\textbf {\bibinfo
  {volume} {5}},\ \bibinfo {pages} {13197} (\bibinfo {year}
  {2015})}\BibitemShut {NoStop}%
\bibitem [{\citenamefont {Wen}\ \emph {et~al.}(2016)\citenamefont {Wen},
  \citenamefont {Chen}, \citenamefont {Zhang}, \citenamefont {Chen},\ and\
  \citenamefont {Yu}}]{PhysRevA.94.013829}%
  \BibitemOpen
  \bibfield  {author} {\bibinfo {author} {\bibfnamefont {Y.}~\bibnamefont
  {Wen}}, \bibinfo {author} {\bibfnamefont {Y.}~\bibnamefont {Chen}}, \bibinfo
  {author} {\bibfnamefont {Y.}~\bibnamefont {Zhang}}, \bibinfo {author}
  {\bibfnamefont {H.}~\bibnamefont {Chen}}, \ and\ \bibinfo {author}
  {\bibfnamefont {S.}~\bibnamefont {Yu}},\ }\Doi {10.1103/PhysRevA.94.013829}
  {\bibfield  {journal} {\bibinfo  {journal} {Phys. Rev. A},\ }\textbf
  {\bibinfo {volume} {94}},\ \bibinfo {pages} {013829} (\bibinfo {year}
  {2016})}\BibitemShut {NoStop}%
\bibitem [{\citenamefont {Bastiaans}(1979)}]{Bastiaans:79}%
  \BibitemOpen
  \bibfield  {author} {\bibinfo {author} {\bibfnamefont {M.~J.}\ \bibnamefont
  {Bastiaans}},\ }\Doi {10.1364/JOSA.69.001710} {\bibfield  {journal} {\bibinfo
   {journal} {J. Opt. Soc. Am.},\ }\textbf {\bibinfo {volume} {69}},\ \bibinfo
  {pages} {1710} (\bibinfo {year} {1979})}\BibitemShut {NoStop}%
\bibitem [{\citenamefont {Waller}\ \emph {et~al.}(2012)\citenamefont {Waller},
  \citenamefont {Situ},\ and\ \citenamefont {Fleischer}}]{Nat.Photon.2012}%
  \BibitemOpen
  \bibfield  {author} {\bibinfo {author} {\bibfnamefont {L.}~\bibnamefont
  {Waller}}, \bibinfo {author} {\bibfnamefont {G.}~\bibnamefont {Situ}}, \ and\
  \bibinfo {author} {\bibfnamefont {J.~W.}\ \bibnamefont {Fleischer}},\ }\Doi
  {10.1038/nphoton.2012.144} {\bibfield  {journal} {\bibinfo  {journal} {Nat.
  Photon.},\ }\textbf {\bibinfo {volume} {6}},\ \bibinfo {pages} {474}
  (\bibinfo {year} {2012})}\BibitemShut {NoStop}%
\bibitem [{\citenamefont {Alonzo}\ \emph {et~al.}(2005)\citenamefont {Alonzo},
  \citenamefont {Rodrigo},\ and\ \citenamefont {Gl\"{u}ckstad}}]{Alonzo:05}%
  \BibitemOpen
  \bibfield  {author} {\bibinfo {author} {\bibfnamefont {C.~A.}\ \bibnamefont
  {Alonzo}}, \bibinfo {author} {\bibfnamefont {P.~J.}\ \bibnamefont {Rodrigo}},
  \ and\ \bibinfo {author} {\bibfnamefont {J.}~\bibnamefont {Gl\"{u}ckstad}},\
  }\Doi {10.1364/OPEX.13.001749} {\bibfield  {journal} {\bibinfo  {journal}
  {Opt. Express},\ }\textbf {\bibinfo {volume} {13}},\ \bibinfo {pages} {1749}
  (\bibinfo {year} {2005})}\BibitemShut {NoStop}%
\bibitem [{\citenamefont {Daria}\ \emph {et~al.}(2011)\citenamefont {Daria},
  \citenamefont {Palima},\ and\ \citenamefont {Gl\"{u}ckstad}}]{Daria:11}%
  \BibitemOpen
  \bibfield  {author} {\bibinfo {author} {\bibfnamefont {V.~R.}\ \bibnamefont
  {Daria}}, \bibinfo {author} {\bibfnamefont {D.~Z.}\ \bibnamefont {Palima}}, \
  and\ \bibinfo {author} {\bibfnamefont {J.}~\bibnamefont {Gl\"{u}ckstad}},\
  }\Doi {10.1364/OE.19.000476} {\bibfield  {journal} {\bibinfo  {journal} {Opt.
  Express},\ }\textbf {\bibinfo {volume} {19}},\ \bibinfo {pages} {476}
  (\bibinfo {year} {2011})}\BibitemShut {NoStop}%
\bibitem [{\citenamefont {Liu}\ \emph {et~al.}(2014)\citenamefont {Liu},
  \citenamefont {Mehmood}, \citenamefont {Huang}, \citenamefont {Ke},
  \citenamefont {Ye}, \citenamefont {Genevet}, \citenamefont {Zhang},
  \citenamefont {Danner}, \citenamefont {Yeo}, \citenamefont {Qiu},\ and\
  \citenamefont {Teng}}]{ADOM:ADOM201400315}%
  \BibitemOpen
  \bibfield  {author} {\bibinfo {author} {\bibfnamefont {H.}~\bibnamefont
  {Liu}}, \bibinfo {author} {\bibfnamefont {M.~Q.}\ \bibnamefont {Mehmood}},
  \bibinfo {author} {\bibfnamefont {K.}~\bibnamefont {Huang}}, \bibinfo
  {author} {\bibfnamefont {L.}~\bibnamefont {Ke}}, \bibinfo {author}
  {\bibfnamefont {H.}~\bibnamefont {Ye}}, \bibinfo {author} {\bibfnamefont
  {P.}~\bibnamefont {Genevet}}, \bibinfo {author} {\bibfnamefont
  {M.}~\bibnamefont {Zhang}}, \bibinfo {author} {\bibfnamefont
  {A.}~\bibnamefont {Danner}}, \bibinfo {author} {\bibfnamefont {S.~P.}\
  \bibnamefont {Yeo}}, \bibinfo {author} {\bibfnamefont {C.-W.}\ \bibnamefont
  {Qiu}}, \ and\ \bibinfo {author} {\bibfnamefont {J.}~\bibnamefont {Teng}},\
  }\Doi {10.1002/adom.201400315} {\bibfield  {journal} {\bibinfo  {journal}
  {Adv. Opt. Mater.},\ }\textbf {\bibinfo {volume} {2}},\ \bibinfo {pages}
  {1193} (\bibinfo {year} {2014})}\BibitemShut {NoStop}%
\bibitem [{\citenamefont {Schley}\ \emph {et~al.}(2014)\citenamefont {Schley},
  \citenamefont {Kaminer}, \citenamefont {Greenfield}, \citenamefont
  {Bekenstein}, \citenamefont {Lumer},\ and\ \citenamefont
  {Segev}}]{schley2014loss}%
  \BibitemOpen
\bibfield  {journal} {  }\bibfield  {author} {\bibinfo {author} {\bibfnamefont
  {R.}~\bibnamefont {Schley}}, \bibinfo {author} {\bibfnamefont
  {I.}~\bibnamefont {Kaminer}}, \bibinfo {author} {\bibfnamefont
  {E.}~\bibnamefont {Greenfield}}, \bibinfo {author} {\bibfnamefont
  {R.}~\bibnamefont {Bekenstein}}, \bibinfo {author} {\bibfnamefont
  {Y.}~\bibnamefont {Lumer}}, \ and\ \bibinfo {author} {\bibfnamefont
  {M.}~\bibnamefont {Segev}},\ }\href {http://dx.doi.org/10.1038/ncomms6189}
  {\bibfield  {journal} {\bibinfo  {journal} {Nat. Commun.},\ }\textbf
  {\bibinfo {volume} {5}},\ \bibinfo {pages} {5189} (\bibinfo {year}
  {2014})}\BibitemShut {NoStop}%
\bibitem [{\citenamefont {L\'opez-Aguayo}\ \emph {et~al.}(2010)\citenamefont
  {L\'opez-Aguayo}, \citenamefont {Kartashov}, \citenamefont {Vysloukh},\ and\
  \citenamefont {Torner}}]{PhysRevLett.105.013902}%
  \BibitemOpen
  \bibfield  {author} {\bibinfo {author} {\bibfnamefont {S.}~\bibnamefont
  {L\'opez-Aguayo}}, \bibinfo {author} {\bibfnamefont {Y.~V.}\ \bibnamefont
  {Kartashov}}, \bibinfo {author} {\bibfnamefont {V.~A.}\ \bibnamefont
  {Vysloukh}}, \ and\ \bibinfo {author} {\bibfnamefont {L.}~\bibnamefont
  {Torner}},\ }\Doi {10.1103/PhysRevLett.105.013902} {\bibfield  {journal}
  {\bibinfo  {journal} {Phys. Rev. Lett.},\ }\textbf {\bibinfo {volume}
  {105}},\ \bibinfo {pages} {013902} (\bibinfo {year} {2010})}\BibitemShut
  {NoStop}%
\end{thebibliography}

%

\end{document}